\documentclass{aastex631}
\pdfoutput=1 
\usepackage{natbib}
\let\tablenum\relax
\usepackage{siunitx}
\usepackage{chemformula}
\usepackage{gensymb}
\usepackage{amssymb}
\usepackage{subcaption}
\usepackage{graphicx}
\usepackage[super]{nth}
\usepackage[version=4]{mhchem}

\DeclareSIUnit\wn{\cm\tothe{-1}}
\DeclareSIUnit\xs{\cm\tothe{-2}}

\DeclareSIUnit\um{\micro\metre}

\begin{document}

\received{December 13, 2021}
\revised{April 3, 2023}
\accepted{April 28, 2023}

\title{Spitzer IRS Observations of Titan as a Precursor to JWST MIRI Observations}

\author{Brandon Park Coy}
\altaffiliation{Now at the Department of Earth, Planetary, and Space Sciences, UCLA, Los Angeles, California, United States}
\affiliation{Center for Research and Exploration in Space Science \& Technology II (CRESST II)}
\affiliation{NASA Goddard Space Flight Center, Greenbelt, Maryland, United States}

\affiliation{Southeastern Universities Research Association (SURA)}

\email{bpcoy0@g.ucla.edu}

\author{Conor A. Nixon}
\affiliation{NASA Goddard Space Flight Center, Greenbelt, Maryland, United States}

\author{Naomi Rowe-Gurney}

\affiliation{NASA Goddard Space Flight Center, Greenbelt, Maryland, United States}
\affiliation{Department of Astronomy, University of Maryland, College Park, Maryland, United States}

\affiliation{Center for Research and Exploration in Space Science \& Technology II (CRESST II)}

\affiliation{School of Physics and Astronomy, University of Leicester, Leicester, United Kingdom}

\author{Richard Achterberg}
\affiliation{Center for Research and Exploration in Space Science \& Technology II (CRESST II)}
\affiliation{NASA Goddard Space Flight Center, Greenbelt, Maryland, United States}
\affiliation{Department of Astronomy, University of Maryland, College Park, Maryland, United States}

\author{Nicholas A. Lombardo}
\affiliation{Department of Earth and Planetary Sciences, Yale University, New Haven, Connecticut, United States}

\author{Leigh N. Fletcher}
\affiliation{School of Physics and Astronomy, University of Leicester, Leicester, United Kingdom}

\author{Patrick Irwin}
\affiliation{Department of Physics, University of Oxford, Oxford, United Kingdom}

\begin{abstract}
 In this work we present, for the first time, infrared spectra of Titan from the Spitzer Space Telescope ($2004-2009$).  The data are from both the short wavelength-low resolution (SL, $5.13-\SI{14.29}{\micro\metre}, R\sim60-127$) and short wavelength-high resolution channels (SH, $9.89 - \SI{19.51}{ \micro\metre}, R\sim600$) showing the emissions of \ch{CH4}, \ch{C2H2}, \ch{C2H4}, \ch{C2H6}, \ch{C3H4}, \ch{C3H6}, \ch{C3H8}, \ch{C4H2}, \ch{HCN}, \ch{HC3N}, and \ch{CO2}. We compare the results obtained for Titan from Spitzer to those of the Cassini Composite Infrared Spectrometer (CIRS) for the same time period, focusing on the $16.35-\SI{19.35}{\micro\metre}$ wavelength range observed by the SH channel but impacted by higher noise levels in CIRS observations. We use the SH data to provide estimated haze extinction cross-sections for the $16.67-\SI{17.54}{\um}$ range that are missing in previous studies. We conclude by identifying spectral features in the $16.35-\SI{19.35}{\micro\metre}$ wavelength range, including two prominent emission features at 16.39 and \SI{17.35}{\micro\metre}, that could be analyzed further through upcoming James Webb Space Telescope Cycle 1 observations with the Mid-Infrared Instrument ($5.0-\SI{28.3}{\micro\metre}, R\sim1500-3500$). We also highlight gaps in current spectroscopic knowledge of molecular bands, including candidate trace species such as \ch{C60} and detected trace species such as \ch{C3H6}, that could be addressed by theoretical and laboratory study. 
\end{abstract}

\section{Introduction}  

Saturn’s moon Titan exhibits the most complex and diverse atmospheric chemistry of any body in the solar system besides Earth. The bulk atmospheric composition has been estimated at mole fractions of $1.5-5.5\%$ methane (\ch{CH4}) and $98.5-94.5\%$ nitrogen gas (\ch{N2}) depending on altitude \citep{Niemann2010}. Photochemistry in the upper atmosphere produces a rich array of hydrocarbons and nitriles via dissociation of methane and nitrogen gas \citep{Lavvas2008,Loisin2015}. Some oxygen compounds including carbon monoxide (\ch{CO}), carbon dioxide (\ch{CO2}), and water vapor (\ch{H2O}) are also produced due to an external oxygen source from Enceladus and contribute heavily to various chemical processes in the atmosphere \citep{horst2008,Dobrijevic2014,Teanby2018,Vuitton2019}.

In addition, Titan contains a complex organic haze with a vertical density profile and albedo wavelength dependence that is still not well-constrained \citep{Li2011,li2015,creecy2019}.  The haze, primarily composed of organic-rich solid particles, absorbs at visible wavelengths and is transparent at infrared (IR) wavelengths. This leads to an anti-greenhouse effect in which solar radiation is absorbed in the upper atmosphere whereas thermal IR radiation from the surface escapes easily, reducing the surface temperature by an estimated 9 K \citep{Mckay1991}.

Since the 1970s, IR spectroscopy has been a useful tool for probing the composition of the neutral atmosphere of Titan.  This has been conducted using ground-based telescopes, visiting spacecraft such as Voyager and Cassini, and space-based observatories including theInfrared Space Observatories.  Data from the Voyager 1 Infrared Interferfometer Spectrometer and Radiometer (IRIS) provided moderate-resolution observations to identify previously-undiscovered molecules, including diacetylene (\ch{C4H2}), cyanoacetylene (\ch{HC3N}), cyanogen [\ch{(CN)2}], propane (\ch{C3H8}), and propyne \citep[\ch{CH3CCH} or \ch{C3H4},][]{Kunde1981,maguire1981}. The Infrared Space Observatory (ISO) Short Wavelength Spectrometer (SWS) offered a resolving power $5-20$ times higher than IRIS and led to the discovery of water vapor emission bands near \SI{40}{\micro\metre} and benzene (\ch{C6H6}) at \SI{14.8}{\micro\metre} \citep{Coustenis1998,Coustenis2003}. Data from Cassini CIRS were able to map the evolving spatial distribution of molecules as Titan progressed through half its year from 2004 through 2017.  However, given that the spectral resolution was comparable to or lower than that of ISO, it did little to provide detection of new molecular species, with propylene (\ch{C3H6}) being the only new molecule discovered on Titan with CIRS \citep{Nixon_2013}.  Propadiene (\ch{CH2CCH2}), an isomer of propyne, has also recently been detected using data from the ground-based TEXES instrument at the NASA Infrared Telescope Facility \citep{Lombardo2019c}.

The Spitzer Space Telescope \citep{Werner2004} was the third spacecraft dedicated to IR astronomy after the Infrared Astronomical Satellite and ISO.  Spitzer's main telescope contained three instruments, the Infrared Array Camera \citep[IRAC,][]{fazio2004}, Infrared Spectrograph \citep[IRS,][]{Houck2004a}, and Multiband Imaging Photometer for Spitzer \citep[MIPS,]{rieke2004}. Spitzer IRS data has been used to measure the atmospheric compositions of many other solar system bodies, including Uranus \citep{BURGDORF2006,ORTON2014a,ORTON2014b,Rowe_Gurney_2021} and Neptune \citep{Meadows2008}.  Although Spitzer performed multiple dedicated observations of Titan, mostly for calibration purposes, none of the data have been modeled or published before.

IRS observations largely overlap with those of Cassini CIRS -- both focused on a similar wavelength range and overlap in seasonal coverage. While Cassini CIRS provided a vast amount of information on Titan's atmosphere, Cassini CIRS detector Focal Plane 1 (FP1) exhibited heightened noise levels at the short-wavelength edge of its range \citep[$\sim17-\SI{20}{\micro\metre}$,][]{Jennings:17}. 
IRS SH observations cover a majority of this noisy region ($17.00-\SI{19.51}{\micro\metre}$) and can provide new high spectral resolution coverage in addition to serving as a check on previous CIRS results.

The James Webb Space Telescope (JWST) Mid-Infrared Instrument \citep[MIRI,][]{rieke2015mid} is scheduled to observe Titan and is expected to make several improvements over IRS, including a threefold increase in resolving power in the $17-\SI{20}{\micro\metre}$ range ($R\sim1800$ vs. $R\sim600$) and an over tenfold increase in spatial resolution ($0.20''-0.27''$ per pixel vs. 2.3$''$ per pixel), allowing for partial spatial resolution of Titan's disk (diameter $\sim0.84''$).  Thus, studying the $17-\SI{19.51}{\micro\metre}$ range observed by Spitzer is important for identifying possible spectral features for follow up with MIRI that may have been obstructed by noise in Cassini CIRS observations.  Comparisons of the observational parameters of the Spitzer IRS alongside those of previous IR space-based observations of Titan and JWST instruments are shown in Table \ref{tab:obs_comparisons}.

\begin{deluxetable*}{llllll}
\tablecaption{Observational parameters for select Titan mid-infrared spectroscopic observations from space\label{tab:obs_comparisons}}
\tablewidth{0pt}
\tablehead{
\colhead{Instrument} & \colhead{Exploitable Range ($\mu$m)} &\colhead{Approximate Resolving Power}  & \colhead{Data Timespan (years)}
}
\startdata
Voyager 1 IRIS \tablenotemark{a} & $7.0-50.0$ & $50-500$ & 1980 \\
ISO SWS Grating Mode \tablenotemark{a} & $7.0-45.0$ & $1500-3000$ & 1997 \\
Spitzer IRS SL & $7.0-11.7$ & $60-127$ & $2004-2006$\\
Spitzer IRS SH & $9.9-19.5$ & $600$  &  $2004-2008$ \\
Cassini CIRS FP1 \tablenotemark{b} & $20.0-1000.0$ & $20-1000$ &  $2004-2017$ \\
Cassini CIRS FP3 \tablenotemark{b} & $9.1-16.7$ & $1200-2200$ & $2004-2017$ \\
Cassini CIRS FP4 \tablenotemark{b} & $7.1-9.1$ & $2200-2500$ & $2004-2017$ \\
JWST NIRSpec \tablenotemark{c} & $0.6-5.3$ & $2700$ & $2022~-$ \\
JWST MIRI \tablenotemark{d} & $5.0-28.3$ & $1500-3500$ & Scheduled ($2023~-$)\\
\enddata
\tablecomments{``Exploitable Range" refers to the range of the data that showed a high enough signal-to-noise ratio suitable for scientific analysis.} 
\tablenotetext{a}{\citet{Coustenis2003}}
\tablenotetext{b}{\citet{Nixon2019}}
\tablenotetext{c}{\citet{jakobsen2022}}
\tablenotetext{d}{\citet{labiano2021}}
\end{deluxetable*}

In this work, we summarize our data reduction process and spectral retrieval results using Spitzer IRS spectra of Titan.  Section \ref{sec:obs} describes the observation data format along with the calibration and reduction processes.  Section \ref{sec:retrievals} describes the spectral fitting and retrieval process, with Section \ref{sec:results} discussing our retrieval results. 
Section \ref{sec:discussion} compares our temperature retrievals with CIRS-based retrievals presented in \citet{Teanby2019} and introduces candidate spectral features for follow-up observations with JWST.  Section \ref{sec:conclusion} contains our conclusions and recommendations for future work on Titan.

\section{Observations}
\label{sec:obs}

Spitzer IRS observed Titan in three out of four of its available modules. This includes data from the short wavelength-low resolution modules (SL, $R=60\sim127$, $5.13 - \SI{14.29}{\micro\metre}$), short wavelength-high resolution module (SH, $R\approx600, 9.89 - \SI{19.51}{ \micro\metre}$), and long wavelength-high resolution modules (LH, $R\approx600, 18.83 - \SI{37.14}{ \micro\metre}$). Unlike SL and SH data, flux values in LH exposures were largely inconsistent over separate observations and not used in this analysis (see Appendix). The diffraction grating of each module splits them into smaller wavelength range `orders'. The SH module contains 10 separate orders and the SL module contains 3 (SL1, SL2, and SL3). While SH order data are taken simultaneously, SL data collection is split into two steps (see Section \ref{sec:overlap}).  Table \ref{tab:modules} lists various observational parameters for each module.

The $\sim 0.84''$ angular diameter disk of Titan was much smaller than the 1.8$''$ and 2.3$''$ per pixel scale of the SL and SH modules respectively and was thus treated as an unresolved source with spectra being considered disk-averaged. Data were taken in both the staring image mode and the mapping image mode (Section \ref{mapping_section}). A portion of the mapping mode data taken by Spitzer were originally intended to be analyzed for the difference in spectral signatures of the hemisphere containing the ``continent" (i.e. Xanadu) seen in near-infrared Hubble images and the trailing hemisphere \citep{Houck2004b}, but these plans never came to fruition.  The remaining Titan data were taken to assist in the calibration of the IRS SH module, and were not originally intended for scientific analysis.

\subsection{Mapping Mode}
\label{mapping_section}

The majority of the data used in our analysis were taken in an IRS imaging mode known as mapping mode. As opposed to staring mode, the standard operating mode of the IRS that centers the object in the module's slit, mapping mode took exposures in a grid of map positions around a central object.  Data taken over a full grid is referred to as a single cycle.  This imaging mode was helpful for extended sources that cover an area of the sky larger than the module's slit dimensions, but when viewed by IRS, Titan is essentially a point source, and thus mapping mode provided few advantages over staring mode. Additionally, when utilizing mapping mode there is not necessarily a grid position where the target is entirely centered on the slit. This means that Titan was not always centered, or even contained within the slit boundaries, leading to a significant drop in brightness in many exposures. Modifications made to the analysis of mapping mode data are discussed in Section \ref{sec:filling_factor}.

Unlike stare data, map exposures use a single center position nod and not two separate \nth{1} position and \nth{2} position nods. This led to a slight modification in the sky subtraction step from the methodology presented in \citet{Rowe_Gurney_2021}, outlined in Section \ref{sec:errors}.

\begin{figure}
\centering
\begin{subfigure}{.4\textwidth}
  \centering
  \includegraphics[width=\linewidth]{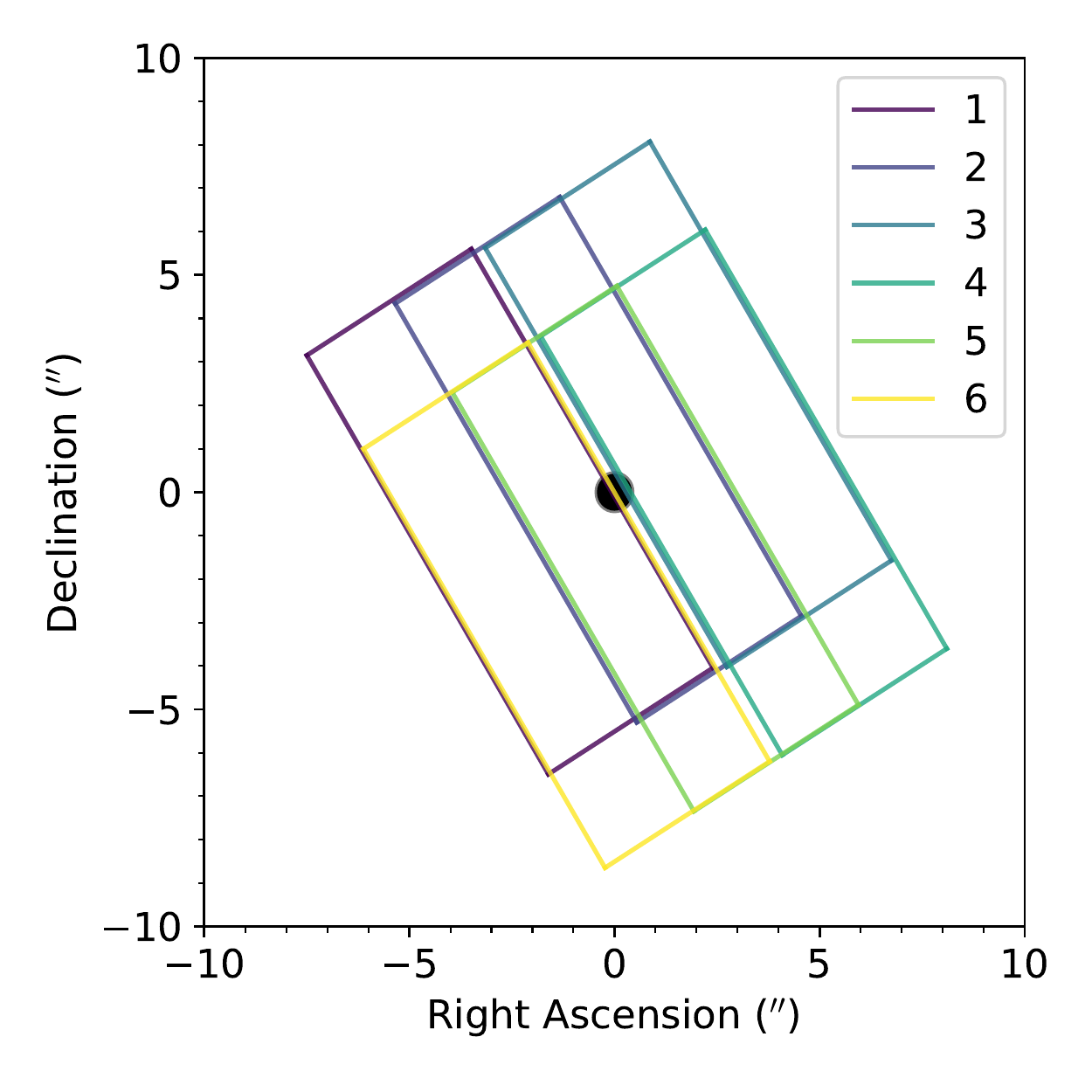}
  \caption{SH module, 2005-11-15, Map Dataset 1}
  \label{fig:sub1}
\end{subfigure}
\begin{subfigure}{.4\textwidth}
  \centering
  \includegraphics[width=\linewidth]{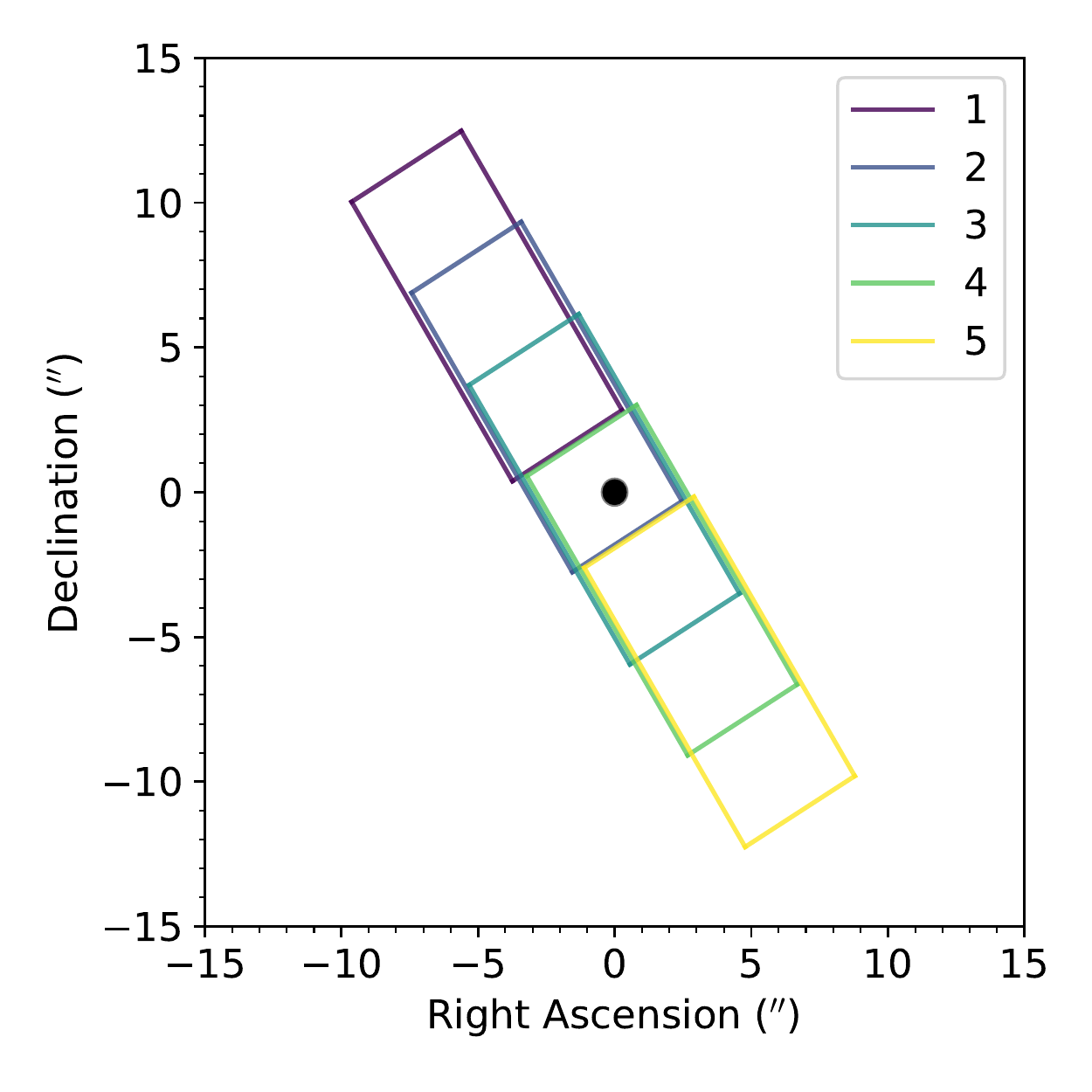}
  \caption{SH module, 2005-11-19, Map Dataset 2}
  \label{fig:sub2}
\end{subfigure}
\caption{Example maps of the area captured by the SH module in (a) both parallel and perpendicular mapping steps and (b) only parallel mapping steps in relation to Titan's position observed by Spitzer (black circle).  Different colors represent different mapping steps for each session.  Right ascension and declination are reported as the deviation from Titan's center. For Map Dataset 1, steps 2 and 5 had filling factors of 1. Titan was partially captured (filling factor between 0 and 1) in steps 1, 3, 4, and 6.  For map dataset 2, steps 2, 3, and 4 had filling factors of 1. Steps 1 and 5 had a filling factor of 0. For Map Dataset 1, SL observations followed the same mapping layout.}
\label{fig:fillingfactor}
\end{figure}

Given that several observations could be grouped based on certain characteristics, we have split the observations into three distinct datasets---Map Dataset 1, Map Dataset 2, and the Stare Dataset.  Observations in Map Dataset 1 contain data from the SL1, SL2, and SH modules and were mapped using a grid of 2 parallel steps and 3 perpendicular steps with a step size of $2.4''$ over 1 mapping cycle. Map Dataset 2 contains data from only the SH module and observations were mapped using a grid of 5 parallel steps with a step size of $3.68''$ over 4 mapping cycles. Example maps for both Map Datasets can be seen in Figure \ref{fig:fillingfactor}. The Stare Dataset contained standard stare exposures of SH over 4 cycles.  Exposure time was constant for each dataset at 6 seconds per exposure for both SL and SH and each module produces $128\times128$ pixel images.

While it spans a larger time-frame than Map Dataset 1, Map Dataset 2 was not used in this analysis as its average flux standard deviation between exposures within the same observation session (7.8\%) was significantly higher than that of Map dataset 1 (3.1\%) and the Stare dataset (2.0\%).  
The combination of Map Dataset 1 (hereafter the Map Dataset) and the Stare Dataset used in this investigation spans a wide time interval of March 4th, 2004 to April 20th, 2006.  
A full list of observation times and various observational parameters is in Table \ref{tab:data}.

\begin{deluxetable*}{llllr}
\tablecaption{IRS Module Parameters\label{tab:modules}}
\tablewidth{0pt}
\tablehead{
\colhead{Module Name} & \colhead{Wavelength Range (\SI{}{\micro\metre})} & \colhead{Resolving Power} & \colhead{Plate Scale ($''$/pixel)} & \colhead{Slit Dimensions ($''$)}
}
\startdata
    SL2 & $5.13-7.60$ & $60-127$ & 1.8 & $(\sim3.6-3.7) \times 57.0$\\
    SL3 & $7.33-8.66$ & $61-120$ & 1.8 & $(\sim3.6-3.7) \times 57.0$ \\
    SL1 & $7.46-14.29$ & $61-120$ & 1.8 & $(\sim3.6-3.7) \times 57.0$ \\
    SH  & $9.89-19.51$ & 600 & 2.3 & $4.7 \times 11.3$ \\
\enddata
\tablecomments{Despite being reported as constant for the SH module, resolving power varies slightly with wavelength \citep{ORTON2014a} and this is an approximate value.}
\end{deluxetable*}

\subsubsection{Filling Factor}
\label{sec:filling_factor}
Due to problems related to mapping mode, we only used exposures where Titan was completely contained within the slit boundaries.  This was done by approximating declination and right ascension as Cartesian x-y coordinates and calculating the rectangular area encased by the slit boundaries using the specific slit's (see Table \ref{tab:modules}) dimensions and pointing angle.  Titan was then approximated as a circle in the right ascension-declination plane, with Titan's angular diameter and sky coordinates (relative to Spitzer at the given observation time) retrieved from the Jet Propulsion Laboratory's Horizons \href{https://ssd.jpl.nasa.gov/horizons/}{online solar system data and ephemeris computation service }\citep{Giorgini1996}.  The full atmosphere of Titan was accounted for by a 425 km (16.5\%) increase in its reported diameter, in accordance with the limb extension radius used in \citet{Teanby2013}. We assumed that their limb extension radius, derived for Herschel data taken over $32-51 \textrm{ cm}^{-1}$ ($196-\SI{312}{\micro\metre}$), is roughly equal to the limb extension radius for our observed wavelengths. A filling factor, the percentage of Titan enclosed within the approximated slit's boundaries, was then calculated for each exposure.   
Only exposures with a filling factor of 1 were considered in this work, which accounted for a third of exposures for Map Dataset 1.
Spectra from individual exposures of an observation were then averaged to help remove noise.

\begin{deluxetable*}{llllllll}
\tablecaption{Spitzer IRS Titan Observations\label{tab:data}}
\tablewidth{0pt}
\tablehead{
\colhead{Date} & \colhead{AORKEY} & \colhead{Start} & \colhead{End Time} & \colhead{$\delta_{Titan}$ ($''$)}  & \colhead{$\theta_{sep}$ ($''$)} & \colhead{\# Filled Exposures} & \colhead{Model Fit RMS Error (\%)}
}
\startdata
 \multicolumn{8}{c}{\textbf{Map Dataset (SL, SH)}} \\
 \hline
2004-03-05 & 4537856 & 07:10 & 07:31 & 0.8303 & 191 & 2 SL \& 2 SH & 2.9 (SL), 6.8 (SH)\\
2004-03-25 & 4538368 & 15:37 & 15:57 & 0.7992 & 86 & 2 \& 2 & 5.5 (SL), 6.4 (SH)\\
2005-11-15 & 4537600 & 03:16 & 03:26 & 0.7817 & 52 & 2 \& 2 & 6.0 (SL), 7.0 (SH)\\
2006-04-20 & 4538112 & 19:48 & 19:59 & 0.8178 & 122 & 2 \& 2 & 3.8 (SL), 6.8 (SH)\\
  \hline
 \multicolumn{8}{c}{\textbf{Stare Dataset (SH only)}} \\
 \hline
2004-03-04 & 9108992 & 19:31 & 19:41 & 0.8311 & 185 & 8 & 5.3\\
2004-11-16 & 12606464 & 16:06 & 16:08 & 0.8200 & 191 & 8 & 5.6\\
2005-03-23 & 13360128 & 22:43 & 23:01 & 0.8311 & 193 & 8 & 5.0\\
2005-04-16 & 13530112 & 21:58 & 22:11 & 0.7956 & 192 & 8 & 4.9\\
\enddata
\tablecomments{Times shown are in UTC. The angular diameter of Titan, $\delta_{Titan}$, and angular separation of Titan and Saturn, $\theta_{sep}$, were calculated using the JPL Horizons online interface.  Exposures from each observation date were averaged to create one spectrum per observation.}
\end{deluxetable*}

Given that Titan is one of the brightest sources viewed by IRS, the SL1 data contained sections of the spectra $\lambda \gtrsim \SI{11.7}{\micro\metre}$ that exhibited signs of oversaturation that had not been automatically detected by the IRS Data Pipeline. This was identified by eye through large mismatches with the SH data flux in the overlap ($\lambda=9.89-\SI{14.29}{\micro\metre}$) region. This region was not used for our spectral retrievals. The non-saturated regions of the SL images did not show signs of significant stray light from the saturated regions.

\subsection{Spectral Extraction}
Spectral extraction was performed in accordance with \citet{Rowe_Gurney_2021} with modifications due to the change in observation target and imaging mode.  Software used in the reduction process is available from the \href{https://irsa.ipac.caltech.edu/data/SPITZER/docs/dataanalysistools/}{Spitzer Science Center} (SSC).

Data are available from the Spitzer Heritage Archive \citep[SHA,][]{https://doi.org/10.26131/irsa543} in the form of FITS files of the 2D detector images. We used the first level Basic Calibrated Data and cleaned them manually as opposed to using second level Post-Basic Calibrated Data data that were automatically cleaned by the pipeline. SHA data contain masks of historically dead and over saturated pixels. We used the SSC program \texttt{irsclean}'s built-in bad pixel detection method alongside identifying remaining bad pixels by eye.  Bad pixels were selected if they were dead or showed large changes in brightness over multiple exposures.  In cases with multiple exposures per location (Stare Dataset only), the provided software \texttt{coad} was used to co-add the exposures.  The SSC program SPitzer IRS Custom Extractor (\texttt{SPICE}) was run with default settings to extract spectra from the cleaned and co-added detector images.  The SSC program \texttt{stinytim v2.0} accounted for overfilling of the slit by Titan's point-spread function (PSF) for \SI{0.5}{\micro\metre} increments of wavelength.

Unlike staring mode observations, SL1 and SL2 data were not taken simultaneously in mapping mode.  However, each mapped sky position included separate SL1 and SL2 exposures. These exposures were typically time separated by 2-3 minutes. When the telescope is utilizing SL1, the other module (SL2) contains blank sky, and vice versa. Therefore, the opposite module (SL1 for SL2/3 and vice versa) was used for sky subtraction.  Sky subtraction of SH data was also tested, but the Spitzer campaign lacked dedicated off-source pointings. Due to our relatively high signal-to-background-noise ratio, sky subtraction of other dedicated background campaigns of other targets taken in a close time proximity to the SH observations showed negligible improvement and thus the process was forgone. 

The JPL Horizons interface was used to calculate various parameters pertaining to Titan for each observation, including right ascension, declination, angular separation from Saturn, angular diameter, and sub-observer latitude.  The angular diameter was used to convert the native flux density units of SPICE, janskys, to units of radiance (\si{W~cm^{-2}~sr^{-1}~\micro m ^{-1}}). An effective IR radius of Titan including the atmosphere of 3000~km was used based on assumptions made in Section \ref{sec:filling_factor}.

\subsubsection{SL Overlap Region}
\label{sec:overlap}
SL data are split into three orders: SL1 ($7.46-\SI{14.29}{\micro\metre}$), SL2 ($5.13-\SI{7.60}{\micro\metre}$), and SL3 ($7.33-\SI{8.66}{\micro\metre}$).  SL3 data are taken simultaneously with SL2 and is treated as a `bonus order', providing a small overlap region ($7.46-\SI{8.66}{\micro\metre}$) with the time-separated SL1 data intended to be used to normalize the SL2/3 and SL1 segments of the spectrum. Given that SL2/3 is more accurate when compared to standard stars according to IRS self-reported error values, the overlap region was used to scale the consistently higher SL1 data to the SL2/3 flux.  Scale factors ranged from 0.8\% to 10.3\% with an average of 3.6\%, varying over each observation session.  Edges of each order were trimmed and the three orders were then combined and averaged to create one spectrum. SH data contained 10 separate orders. The orders are taken concurrently and do not exhibit large amounts of overlap and thus were only trimmed at each order's wavelength range edge and not scaled.

\subsubsection{Stray Light from Saturn}
Previous observations of Titan have been hindered by stray light reflecting off of Saturn \citep{COUSTENIS2002}.  We accounted for this effect in the low-resolution data through background subtraction of the unused/opposite module.  However, the high-resolution exposures lacked such an unused module. Therefore, it was difficult to assess whether stray light from Saturn had a significant impact on the SH data. The angular separation of Titan's center from Saturn's center in each observation ranged from 52$''$ to 193$''$, compared to Saturn's $\sim21''$ angular radius, including its rings.  Despite a lack of dedicated background subtraction, there was no statistically significant difference between the flux observed at low separation and high separation ($\ll0.1\%$), suggesting that Titan was far enough for stray light to not display a major impact.

\subsection{Errors and Calibration}
\label{sec:errors}  
Errors output by SPICE are not sufficient enough to account for various systematic sources of error.  We used the IRS self-reported error values based on disagreement with standard stars of 4.6\% for SL1 and 2.2\% for SL2/3 (presented in the \href{https://irsa.ipac.caltech.edu/data/SPITZER/docs/irs/irsinstrumenthandbook/IRS_Instrument_Handbook.pdf}{IRS Instrument Handbook}). Data in the SL1 and SL3 overlap area used the estimated errors of SL1 and SL3 added in quadrature (total 5.1\%) to account for additional error introduced by normalization. As the IRS does not self-report a value for SH, the 4\% standard deviation reported in \citet{ORTON2014a} was used.
Average PSF correction factors output by \texttt{stinytim v2.0} accounted for an extra $0.88\%$ error in the wavelength range used for the SL module and $0.43\%$ in the SH. 
These two sources of error were added in quadrature with the SPICE 1-sigma error to produce an average estimated radiance error of 4.9\% in the SL module, 4.0\% in the SH module for the Stare Dataset, and 4.1\% for the Map Dataset.

\section{Spectral Retrievals}
\label{sec:retrievals}
\subsection{Radiative Transfer Model}
\label{sec:radtrans}

Vertical temperature and gas composition profiles were derived using the optimal estimation retrieval algorithm, Nonlinear optimal Estimator for MultispEctral analySIS \citep[NEMESIS,][]{Nemesis}, which has been used extensively for modeling of planetary atmospheres in the infrared, including multiple studies using Cassini CIRS data \citep[e.g.][]{Teanby2006,Teanby2007,Nixon2010,COTTINI2012,Teanby2019,Lombardo2019a}.

Our Titan reference atmosphere contained 99 pressure levels ranging from 1.44 to $10^{-8}$ bar (0 to 782 km) and included a pressure-temperature \textit{a priori} profiles for temperature and gas abundance derived from CIRS data used in \citet{Teanby2009}. 
The gases included in our model are nitrogen (\ch{N2}), hydrogen (\ch{H2}), carbon dioxide (\ch{CO2}), carbon monoxide (\ch{CO}), water vapor (\ch{H2O}), methane (\ch{CH4}), hydrogen cyanide (HCN), cyanoacetylene (\ch{HC3N}), acetylene (\ch{C2H2}), ethylene (\ch{C2H4}), ethane (\ch{C2H6}), propyne (\ch{C3H4}), propane (\ch{C3H8}), diacetylene (\ch{C4H2}), cyanogen  (\ch{[CN]2}), and benzene (\ch{C6H6}).  Propylene (\ch{C3H6}) was also included using a theoretical profile from the photochemical model of \citet{Loison2019}. Major isotopologues of methane ($^{13}$\ch{CH4} and \ch{CH3D}), ethane (\ch{CH3$^{13}$CH3}), acetylene (\ch{C2HD} and \ch{C$^{13}$CH2}), carbon dioxide (\ch{$^{13}$CO2}, \ch{OC$^{18}$O}, and \ch{OC$^{17}$O}), and hydrogen cyanide (\ch{H$^{13}$CN} and \ch{HC$^{15}$N}) were also included.  Methane isotopes were separated from each other, whereas all other isotopes were included with their associated molecules.

Retrievals for each separate Map Dataset observation were split into two sequential steps.  In the first step, we used the $7.35-\SI{9.00}{\micro\metre}$ ($1110-1360$ cm$^{-1}$) region of the SL spectra containing the $\nu_{4}$ \ch{CH4} spectral band centered at  \SI{7.7}{\micro\metre}.  Full profile retrievals were run for temperature and haze density, whereas \ch{CH3D} was fit for a simple scaling factor.  All other gas profiles were kept fixed. The temperature retrieval results were then used as the temperature profile for our second step.  As the Stare Dataset did not have accompanying SL data, we skipped this first step and instead used the temperature profile from our original \textit{a priori} profile.

The second step used the full range of SH data, edge trimmed to $9.96-\SI{19.35}{\micro\metre}$ ($517-\SI{1004}{\wn}$), to fit for gas and haze profiles. Full retrievals were done for the haze and \ch{C2H6} abundances. \ch{C2H2}, 
\ch{C2H4}, \ch{C3H4}, \ch{C3H8}, \ch{C4H2}, \ch{HCN}, \ch{HC3N}, and \ch{CO2} profiles were fit by adjusting a single scaling factor. Surface temperature, which is detectable via a low-opacity window at \SI{19}{\um} \citep{Jennings2009}, was also derived.  All other profiles were kept fixed.

\subsection{\texorpdfstring{$k$}-Distributions}
\label{sec:k}
NEMESIS was run using the correlated-$k$ approximation method \citep{Lacis} instead of the more computationally intensive line-by-line method. Correlated-$k$ tables for each gas were created using both SL and SH wavelength grids reported by IRS with 50 quadrature points to integrate over ($g$-ordinates). Default terrestrial air broadening (similar to \ch{N2}) was used in place of the \ch{H2} foreign broadening used in \citet{Rowe_Gurney_2021}.  $k$-tables were primarily created using high-resolution line lists from the GEISA (Gestion et Etude des Informations Spectroscopiques Atmosphériques) \citep{chedin,HUSSON1992,JACQUINETHUSSON2016} 2020 and HITRAN 2020 \citep[High-Resolution Transmission Molecular Absorption Database,][]{GORDON2021} online line-list databases. HITRAN was used for most of the molecules listed in \ref{sec:radtrans}, with GEISA only being used for propyne (\ch{C3H4}) and benzene (\ch{C6H6}) which  are not included in the HITRAN database.

A laboratory-derived pseudo line list created at NASA JPL was used for \ch{C3H8} \citep{Sung2013} as GEISA does not include data for all spectral bands included in this study \citep{Nixon2009}. A similar pseudo line list was also used for \ch{C3H6} \citep{SUNG2018}, which is not included in the current HITRAN or GEISA databases.  A Gaussian telescope filter function was calculated for each wavelength assuming a constant $R=600$ for the SH module and $R=
60-127$, based on wavelength, for the SL module.

\subsection{Haze Profile}
Initial haze density height profiles were calculated based on findings presented in \citet{Tomasko2008}. Using Huygens probe Descent Imager/Spectral Radiometer (DISR) instrument data, they determined that Titan's haze between 60 and 150 km consists of aggregate particles that are composed, on average, of 3000 monomers of radius \SI{0.05}{\micro\metre}. These aggregate particles have a number density of 5 cm$^{-3}$ at a height of 80 km and latitude of 10\degree S and decrease with a scale height of 65 km.  Their data were constrained to a maximum altitude of 150 km.  We assumed that this scale height remains constant above 150 km and that their densities measured at 10\degree S could be approximated in a global average. We used the spectral dependence of the haze particles' refractive indices derived from CIRS data and presented in \citet{ANDERSON2011} for the $500-\SI{560}{\wn}$ ($17.9-\SI{20}{\micro\metre}$) spectral range and \citet{VINATIER2012} for the $610-\SI{1500}{\wn}$ ($6.7-\SI{16.4}{\micro\metre}$) spectral range.

\subsection{Disk Averaging}
\label{sec:discaverage}

The contribution to the disk-averaged spectrum from each observable latitude depends on the observer's field of view. The contribution of Titan's disk and its limb extension were discretized into 20 points, with 8 concentric circles across the solid body and 13 across the limb. Field of view averaging points used in the creation of our synthetic model spectra were derived from Appendix A in \citet{Teanby2013}.  Their study assumes a solid radius of 2575 km, with a 425 km limb extension of the atmosphere. We used the same twenty discrete atmospheric paths with weights $w$ defined by the equation,
\begin{equation}
\label{eq:weights}
    w_{i}=\frac{x_{i}x_{i+1}-x_{i-1}x_{i}}{r^{2}},
\end{equation}
where $x$ is the offset from the sub-observer point/disk center associated with an emission angle and $r$ is the effective radius. These weights define the spectral contribution from each of the aforementioned concentric circles and sum to one.

\section{Model Fitting and Interpretation}
\label{sec:results}

\subsection{Low-Resolution Retrievals}
\label{sec:sl_retrievals}

\begin{figure}
\centering
  \includegraphics[width=.85\linewidth]{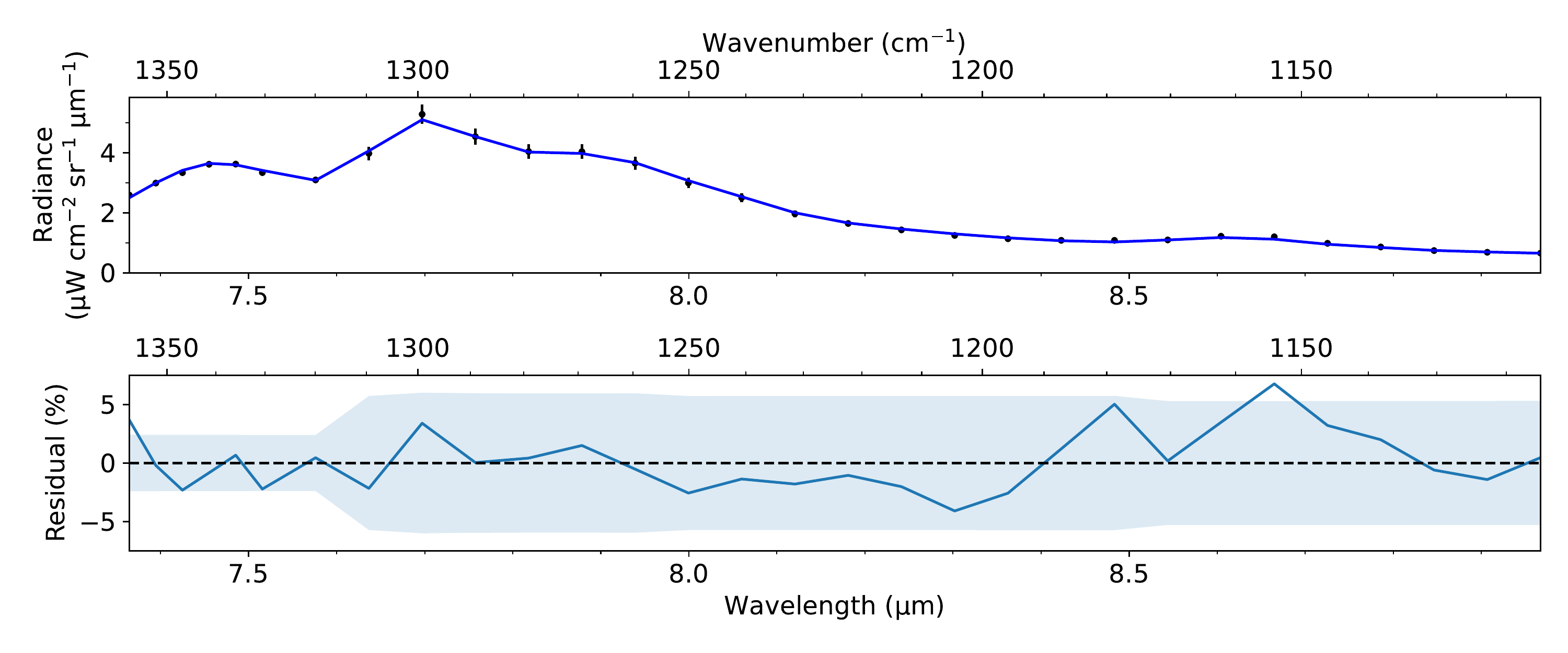}

\caption{Comparison of the Spitzer IRS SL 2004/03/05 spectrum (black dots) of Titan versus our NEMESIS model results (blue line). The total spectrum had a reduced chi-square value $\chi^{2}/N=0.32$. Filled blue regions and black errorbars indicate observational error detailed in Section \ref{sec:errors}.}
    \label{fig:sl_retrieval}
\end{figure}

An example of our SL retrievals is shown in Figure \ref{fig:sl_retrieval}. The methane $\nu_{4}$ band peak, located at \SI{7.7}{\micro\metre}, was consistently weaker in our retrievals compared to observations. This could be due to sources of localized error not accounted for in Section \ref{sec:errors}. 

However, $\chi^{2}/N$ values for some retrievals remain significantly lower than 1, indicating that the error self-reported by IRS may be overestimated, or only applicable for sources much dimmer than Titan. RMS errors for each fit are shown in Table \ref{tab:data}.

\begin{figure}
    \centering
    \includegraphics[width=.8\linewidth]{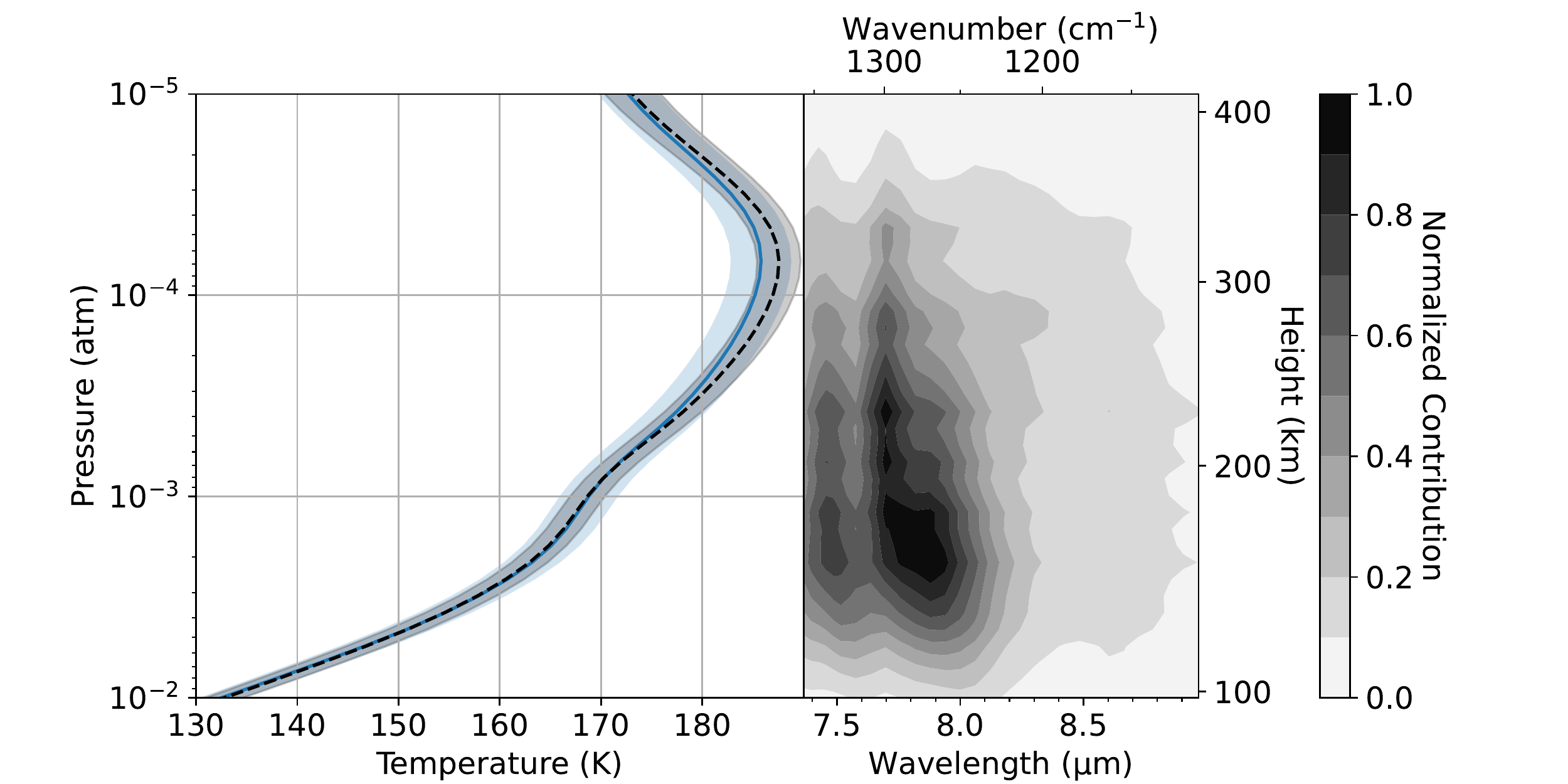}

    \caption{(left) Accompanying temperature retrieval to Figure \ref{fig:sl_retrieval} output by NEMESIS (black dashed line) compared to our \textit{a priori} profile (blue line) alongside the normalized contribution function (right). Filled regions in the retrieved profile indicate assumed error in our \textit{a priori} profile (blue) and fitting error calculated by NEMESIS (black).  Contribution peaks at $\sim2$ mbar (154 km). }
    \label{fig:temp_retrieval}
\end{figure}

Retrieved temperature profiles, an example of which is shown in Figure \ref{fig:temp_retrieval}, exhibited a slight heating in the $250-350$ km height range up to $\sim3$ K compared to our \textit{a priori} profile. Implications of this result are discussed in Section \ref{sec:cirs_temp}. Our calculated contribution functions indicate that our temperature retrievals were mostly sensitive to the $\sim 2$ mbar ($\sim 150$ km) region of the atmosphere, in line with previous temperature retrievals presented in \citet{Teanby2006} and \citet{Coustenis2007}.

\subsection{High-Resolution Retrievals}
\label{sec:sh_retrievals}

\subsubsection{Updated Haze Extinction Cross-Sections}
Extinction cross-section values for Titan's haze were derived from the refractive coefficients presented in \citet{ANDERSON2011} and \citet{VINATIER2012}.  While their data provided coefficients with $\Delta\tilde{\nu}\le\SI{20}{cm^{-1}}$ resolution over the $70-\SI{865} {cm^{-1}}$ ($11.56-\SI{142.86}{\micro\metre}$) range, the data contain a notable \SI{50}{cm^{-1}} gap across $560-\SI{610}{cm^{-1}} (16.39-\SI{17.86}{\micro\metre})$ range, due to higher noise levels in CIRS observations.

Our NEMESIS fits show heightened residuals over the $16.7-\SI{17.1}{\micro\metre}$ range, with a smooth curve indicative of broad haze-like spectral features.  NEMESIS deals with gaps in spectral dependencies by linearly interpolating between points. This is a good approximation if spacing between data points is low, but can lead to inaccurate fitting if there is an extended region with no data. Given that this spectral range will soon be observed with JWST MIRI, providing updated extinction cross-sections for this region will be useful in separating potential underlying spectral features from the broader haze contribution.  

\begin{deluxetable*}{lll}
\tabletypesize{\small}
\tablecaption{Updated haze extinction cross-sections for the $570-\SI{600}{cm^{-1}}$ range. Values in parentheses represent the original interpolated values based on \citet{VINATIER2012}. Uncertainties were calculated based on the standard deviation across our eight observations.\label{tab:haze}}
\tablewidth{0pt}
\tablehead{
\colhead{Wavenumber (\si{\wn})} & \colhead{Wavelength (\si{\um})} & \colhead{Extinction Cross-Section ($10^{-10}\si{\xs}$)}
}
\startdata
570 & 17.544  & $1.664 \pm 0.025$ \hfill (1.631)\\
580 & 17.241  & $1.641 \pm 0.016$ \hfill (1.667)\\
590 & 16.949  & $1.546 \pm 0.034$ \hfill (1.702)\\
600 & 16.667  & $1.668 \pm 0.010$ \hfill (1.736)\\
\enddata
\end{deluxetable*}

\begin{figure}
    \centering
    \includegraphics[width=\linewidth]{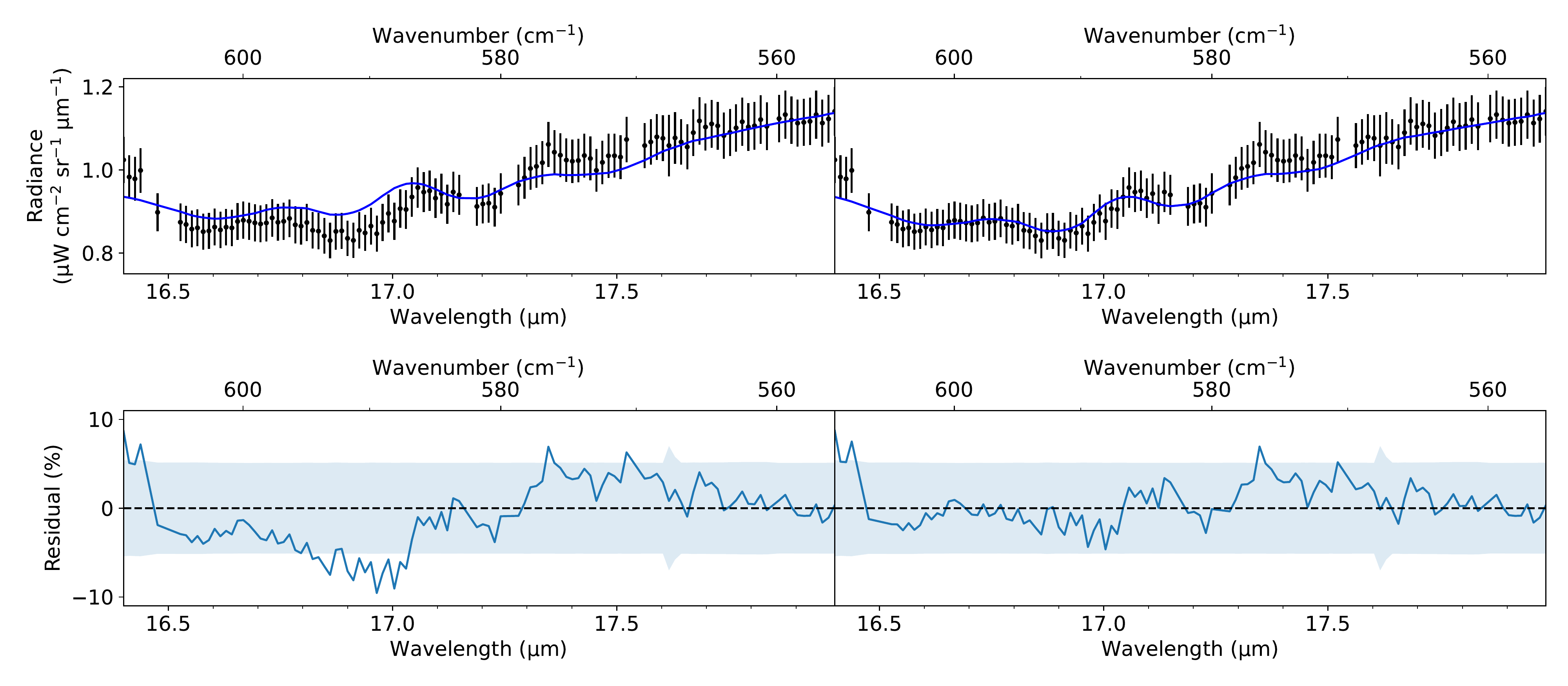}
    \caption{NEMESIS fits for the 2005-11-15 spectrum before (left) and after (right) updating haze extinction cross-sections for the $16.67-\SI{17.54}{\um}$ wavelength range (Table \ref{tab:haze}). The updated fit shows significantly better modelling of the broad haze-like features in the $16.5-\SI{17.1}{\um}$ range, while still retaining narrower signatures of possible molecular emission.}
    \label{fig:updatedhaze}
\end{figure}

We fit for new cross-section values by comparing the results of our NEMESIS fits using the original \citet{VINATIER2012} cross-sections to a gray haze model (equal cross-sections at all wavelengths). The gray haze model was run using the average extinction cross-section of the 560 and $\SI{610}{\wn}$ values, $\sigma=\SI{1.6812e-10}{\xs}$. Residuals for each fit were smoothed using a boxcar moving average to remove random noise. A wide width of 16 wavelength steps ($\sim\SI{0.2}{\micro\metre}$) was used to ensure removal of possible narrow emission features. We masked the prominent features at 16.39 and \SI{17.35}{\micro\metre} that are possibly due to molecular emission.  Difference in residuals at between the gray haze and \citet{VINATIER2012}-derived haze was divided by the difference in their cross sections at  520, 530, 540, 550, 560, and \SI{610}{cm^{-1}} then averaged for an approximate scaling factor.  We then multiplied this scaling factor by our original residuals and added the result to our original cross-sections.  Results can be seen in Table \ref{tab:haze} and Figure \ref{fig:updatedhaze}. Residuals over the $560-\SI{610}{cm^{-1}}$ range decreased by an average of $60\%$. These new extinction cross-sections were then applied to subsequent SH retrievals.

\subsubsection{Wavelength Uncertainties}
\label{sec:shift}
\begin{figure}
    \centering
    \includegraphics[width=.9\linewidth]{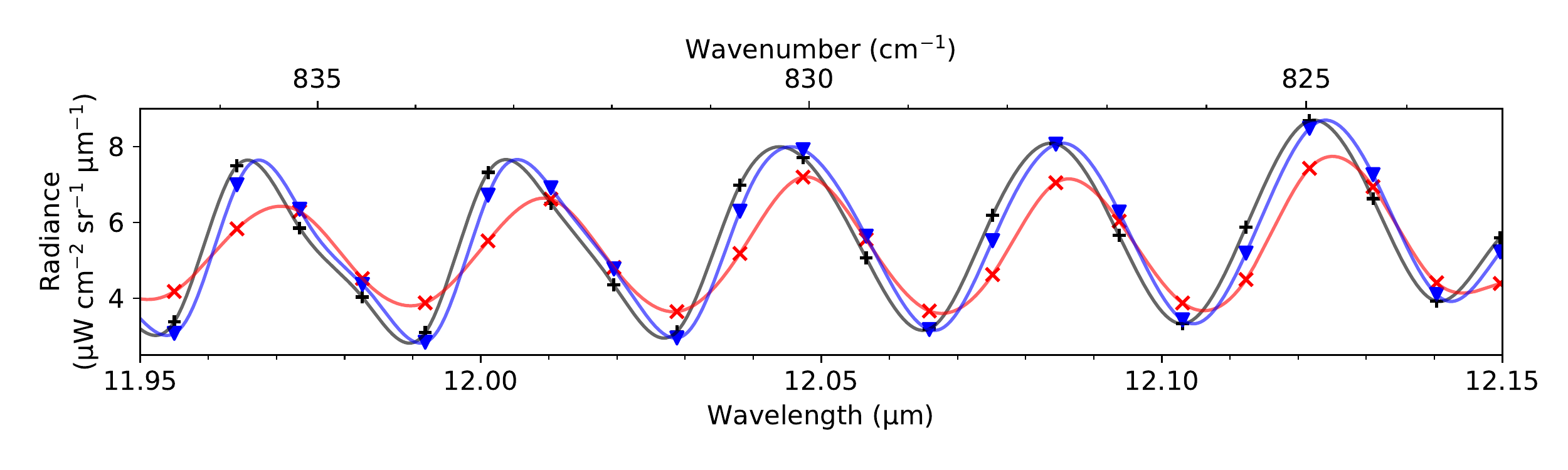}
    \caption{Example results of the cubic spline method described in Section \ref{sec:shift}. The observed data (grey) was cross correlated with our NEMESIS model results (red) to find the wavenumber shift that maximized the cross correlation factor. The shifted spectrum (blue), redshifted by 0.115 cm$^{-1}$, shows noticeably better correlation in spectral band peaks.}
    \label{fig:wavelength_shift}
\end{figure}

After analyzing preliminary SH spectral retrieval results, it became apparent that multiple orders exhibited signs of a slight difference in the wavelength reported by the IRS data pipeline and the true wavelength.  This suggestion is supported by investigations made in \citet{ORTON2014a}, which found that wavelengths can vary by up to $20-25\%$ of a resolution element and found that a wavenumber shift of 0.08 cm$^{-1}$ was required to fit the \ch{H2} S(1) quadrupole spectral feature accurately. Analyses based on CIRS data have required a similar wavenumber shift for accurate model fitting \citep{Nixon2012}.  This wavelength mismatch is likely due to calibration errors in the SH module arising from minute changes in the telescope's internal temperature or incomplete knowledge of the instrument smoothing function.  To counteract this, we fitted both our initial spectral retrievals from NEMESIS and our Spitzer data to cubic splines on a per-order basis.  We then found the wavelength `lag' associated with each order through cross-correlation, restricted to within a fourth of a wavenumber step. We subsequently recalculated our Spitzer cubic spline by shifting our observed wavenumbers by a fitted constant and resampled the new spline for flux values at the original Spitzer wavenumbers.  Lags measured ranged from 0 (most common) to approximately 16.1\% of an wavenumber step.  An example wavenumber shift can be seen in Figure \ref{fig:wavelength_shift}.  This effect is not noticeable in the SL module due to the much lower resolution. The chi-square values in the areas where this fix was applied were significantly reduced, with an average decrease in RMS fitting error of $\sim16\%$ across observations. However, this fix was difficult to apply consistently as it was only calculable for regions with order-wide and well-resolved spectral bands, such as the ethane band centered near \SI{12}{\micro m} and the acetylene band centered near \SI{13.5}{\micro m}.

\subsubsection{Retrieval Results}

\begin{figure}[htp]
\centering
\subfloat[$9.96-\SI{13.09}{\micro\metre}$]{
  \includegraphics[clip,width=.9\linewidth]{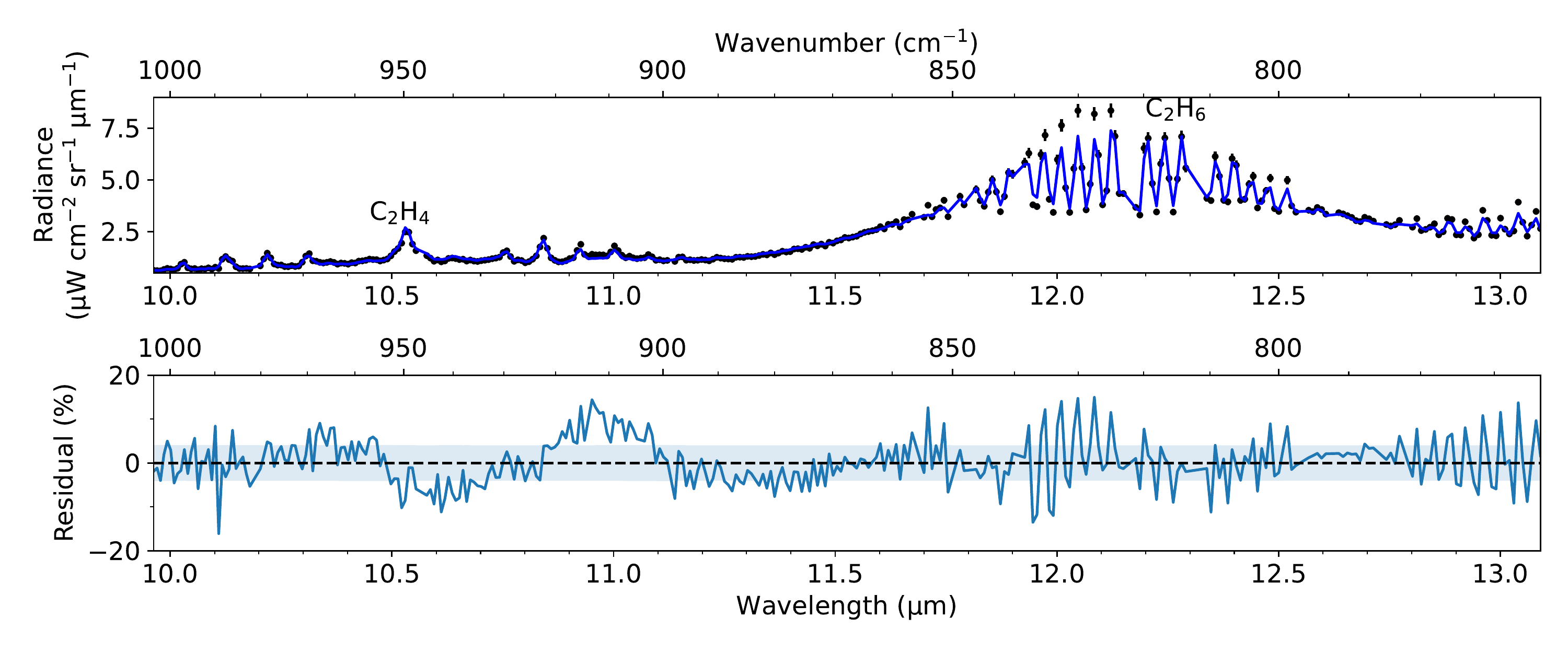}
}

\subfloat[$13.10-\SI{16.21}{\micro\metre}$]{
  \includegraphics[clip,width=0.9\linewidth]{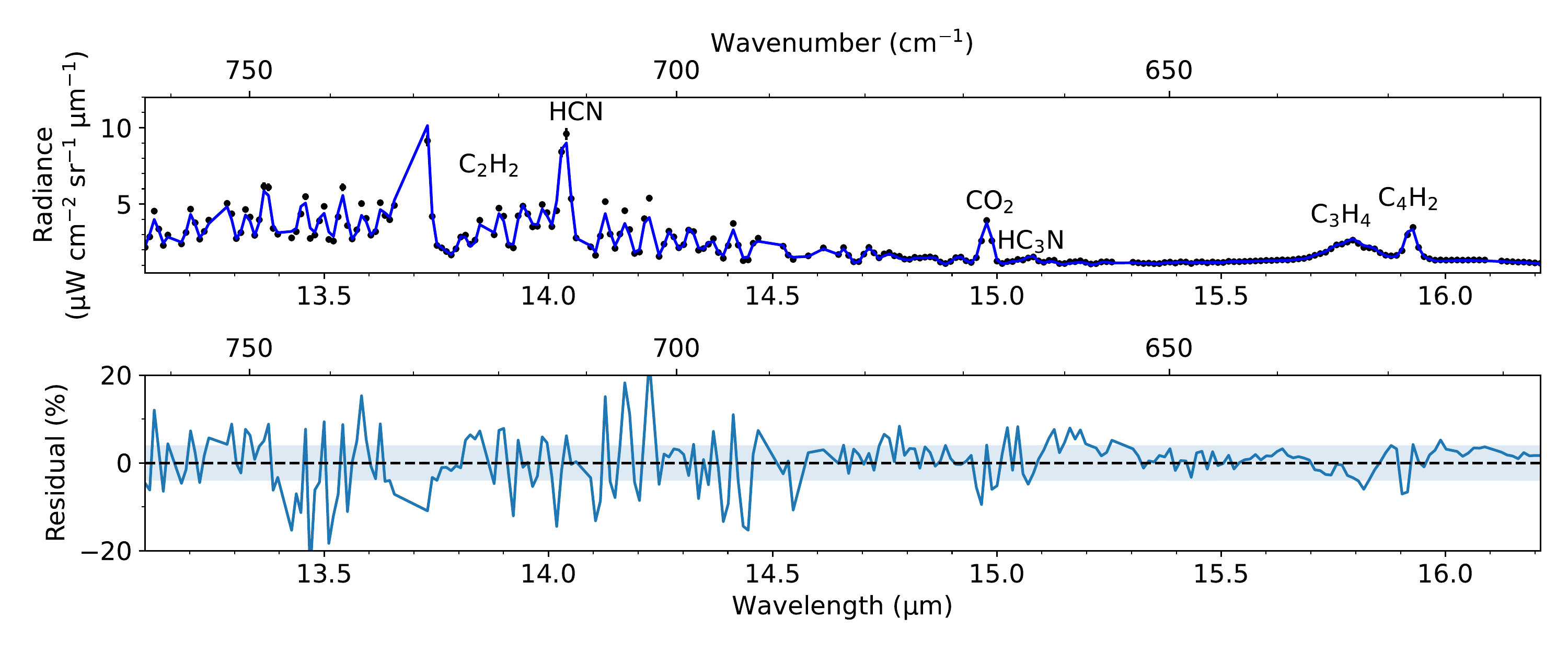}
}

\subfloat[$16.22-\SI{19.35}{\micro\metre}$]{
  \includegraphics[clip,width=0.9\linewidth]{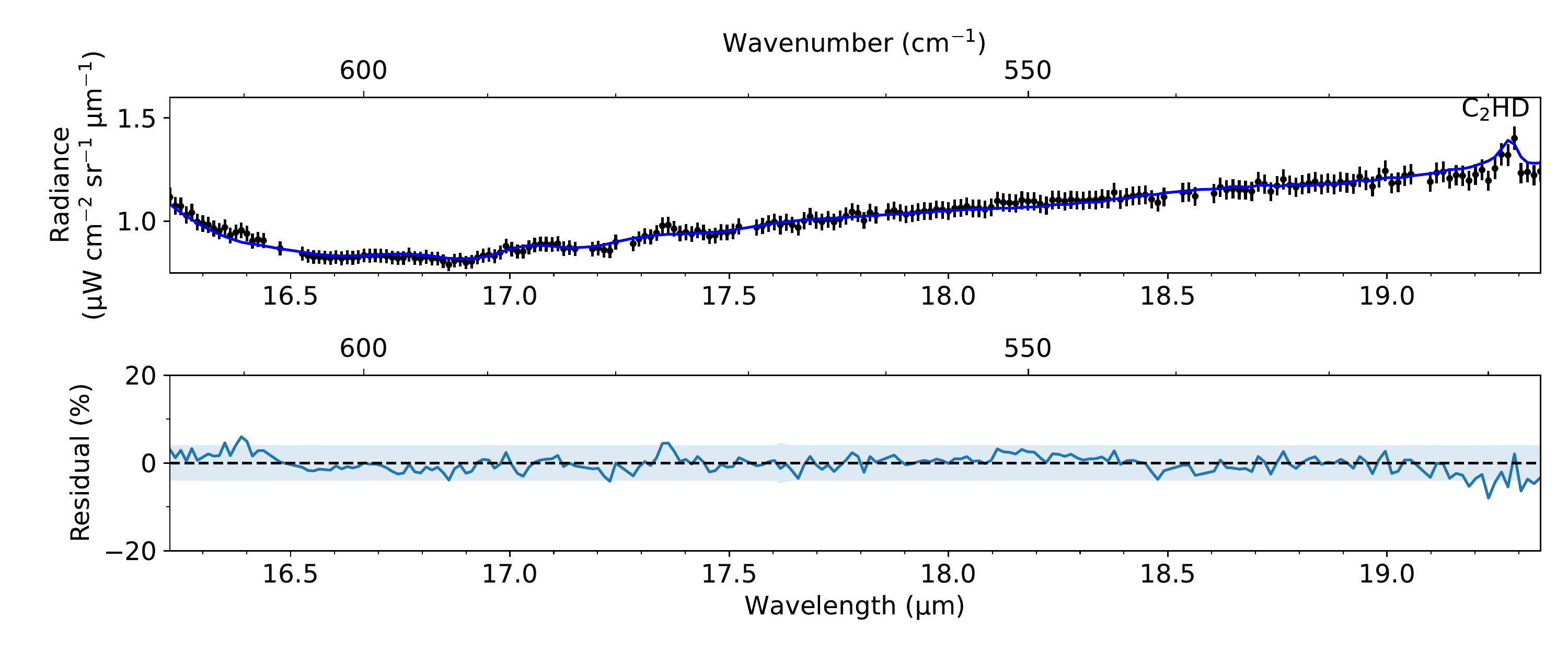}
}
\caption{Comparison of high-resolution 2005/04/16 spectra of Titan versus our NEMESIS model results. The total spectrum had a reduced chi-square value $\chi^{2}/N=1.50$.  Prominent spikes in residual error can be seen in the ethane and acetylene spectral bands as well as band peaks---possible causes are discussed in Section \ref{sec:shift}.
}
    \label{fig:sh_retrieval}
\end{figure}

An example NEMESIS retrieval from the high-resolution data can be seen in Figure \ref{fig:sh_retrieval}.  RMS errors for each fit are shown in Table \ref{tab:data}. The error indicated for the SH retrievals is much higher than seen in Section \ref{sec:sl_retrievals}, particularly in regions with dense spectral bands, discussed below.

\subsubsection{Resolving Power Uncertainties}
Aside from this wavelength shift, peaks and troughs of spectral bands consistently showed residuals above the noise level due to an apparent underestimation by our model (see Figure \ref{fig:sh_retrieval}). Multiple types of filter function shapes for our $k$-table generation (e.g. Lorentzian, triangular, boxcar) were tested but did not show considerable improvement in peak fitting over the Gaussian filter used in \citet{Rowe_Gurney_2021}, in line with results presented in \citet{ORTON2014a}. Multiple types of gas profile models for ethane and acetylene (e.g. simple scaling factor, pressure gradient, full profile retrieval) were also tested but did not show significant improvement. Similar under-fitting of various peaks using NEMESIS can be seen in \citet{Rowe_Gurney_2021}.  Under-fitting of peaks is likely due to a difference between the reported resolving power and the actual resolving power of the telescope.

\begin{figure}

\centering
 \includegraphics[width=\linewidth]{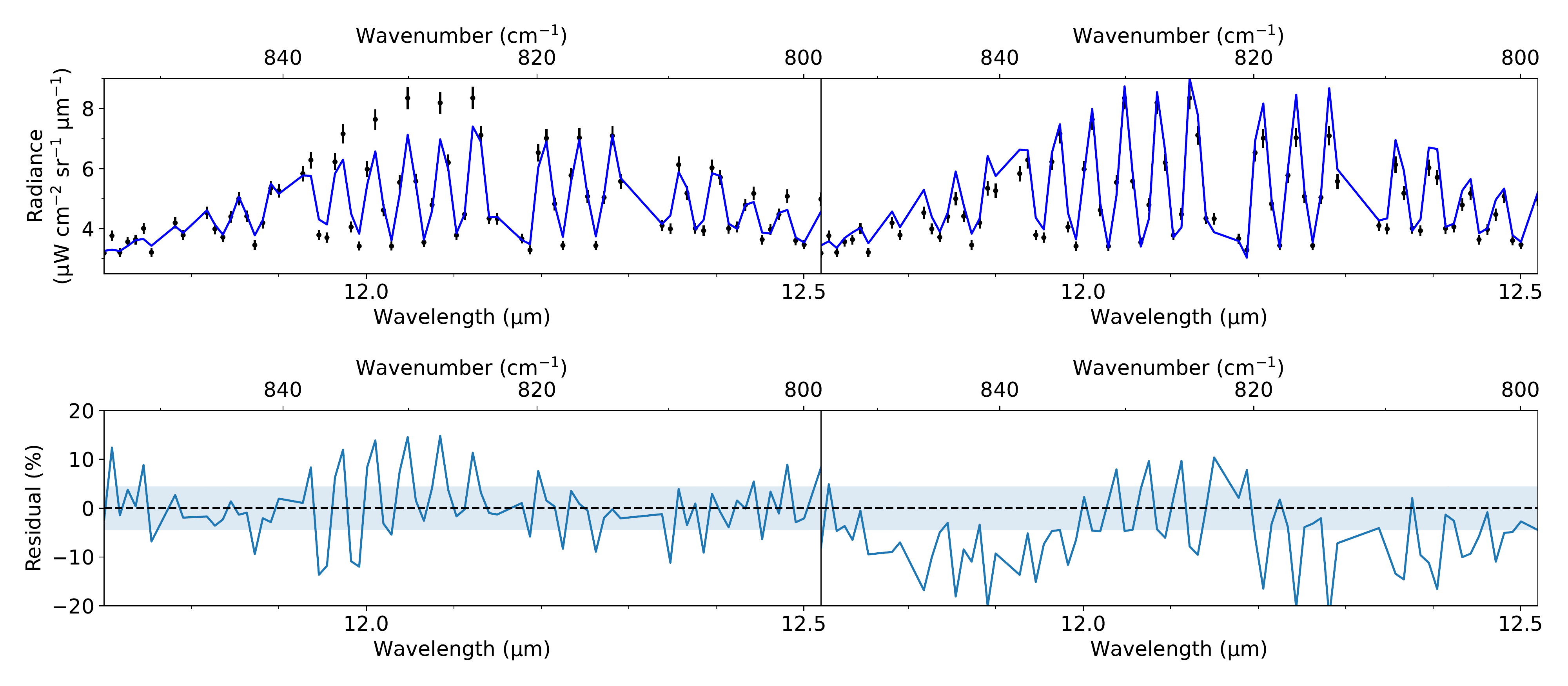}

\caption{Example model fits (blue) of the \ch{C2H6} spectral band near \SI{12}{\micro\metre} assuming a constant resolving power of (left) 600 and (right) 720 compared to Spitzer observations (black).  Higher resolution leads to significantly lower residuals in originally underfit regions ($11.95\sim\SI{12.15}{\micro\metre}$) but higher residuals in originally well-fit regions ($11.75\sim11.95$, $12.15\sim\SI{12.50}{\micro\metre}$). Further studies in dynamical fitting of Spitzer's true resolving power as a function of wavelength will greatly improve radiative transfer model precision.}
\label{fig:resolution}
\end{figure}

Despite being reported as $R\approx600$, the resolving power of the IRS is not constant and varies, with true resolving power $R=600\pm60$ according to the \href{https://irsa.ipac.caltech.edu/data/SPITZER/docs/irs/irsinstrumenthandbook/IRS_Instrument_Handbook.pdf}{IRS Instrument Handbook}. $k$-tables used in our NEMESIS model assume a constant resolving power of $R=600$, and variations in resolving power can greatly affect the shape of our model's spectral features.  Statically increasing the resolution from $R=600$ leads to better fitting in underfit regions of the spectra but overfitting of peaks in already well-fit regions of the spectra (see Figure \ref{fig:resolution}), indicating that the resolution varies significantly as a function of wavelength.

\citet{ORTON2014b} corrected for this effect by radiometrically scaling down SH data to the lower-resolution SL and LL data and fitting a local resolving power on a per-order basis.  They found that for most regions, $R$ only varied as $R=590\sim615$, but more dramatic changes in resolving power were observed if they fit resolving power to specific spectral features, up to $R=743$. Their fix was possible due to data in the overlap region of the SL and SH modules at $9.89-\SI{14.29}{\micro\metre}$ and the LL and SH modules at $13.98-\SI{19.30}{\micro\metre}$. However, as the SL data of Titan are oversaturated at wavelengths $\gtrsim \SI{11.7}{\micro\metre}$, this fix was not applicable to most of our data.
Comparing Spitzer observations of Titan to lower-resolution IR measurements, particularly limb spectra from CIRS, may help in constraining the true IRS resolving power at the ethane and acetylene spectral bands.  

\subsubsection{Gas Profile Retrieval Results} 
\label{sec:shfitting}

Volume mixing ratios for each fitted gas species can be seen in Table \ref{tab:vmrs}. 
We compare our disk averaged measurements to disk averaged profiles from the Seasonally Varying Radiative Species (SVRS) dataset presented in \citet{LOMBARDO2023} \citep[accessible on a Zenodo archive at][]{svrs}.  The SVRS dataset includes abundance information of seven molecules (\ch{C2H6}, \ch{C2H4}, \ch{C2H2}, \ch{C3H4}, \ch{C4H2}, HCN, and \ch{HC3N}) from the surface to an altitude over 500 km (\SI{1e-8}{atm}), across all latitudes and over a full Titan year.  SVRS was developed by combining the meridional and seasonal information of the abundance of each molecule at the \SI{1e-3}{atm} level (about 200 km height) measured with CIRS in \cite{Teanby2019} with the vertical and seasonal information in \cite{Mathe2020}---a complete description may be found in \cite{LOMBARDO2023}.  The interpolation scheme used in SVRS enables a the calculation of a robust disk averaged profile for each molecule, which would otherwise be impossible to determine from the discrete CIRS measurements alone.  In our comparison, we calculate the disk averaged SVRS profiles corresponding to January 2006 (L$_{S}=313\degree$) by a weighted sum of abundance profiles across all latitudes.  The 1-$\sigma$ uncertainties reported here are the weighted 1-$\sigma$ range of abundances across all latitudes.  

Our measurements show a greater abundance in most species than expected from the SRVS derived disk-averaged profiles.  This may be caused by non-linearities in the radiative transfer scheme that are not accounted for in our simplified disk-averaging calculations, especially due to enrichment of trace species near Titan's winter pole.  This enrichment effect has been studied in detail \citep{hourdin2004,Teanby2009,teanby2017,Teanby2019,THELEN2019} and is driven by Titan's stratospheric circulation.  A possible inverse relationship between photochemical lifetimes and enrichment factors at the poles has also been proposed \citep{TEANBY2008}.  This would possibly explain the greater enhancement in our measured \ch{HC3N} compared to the SVRS derived profiles, as \ch{HC3N} has a short photochemical lifetime ($\tau<\SI{1}{year}$) compared to other hydrocarbon and nitrile species \citep{Teanby2006}.

An additional source of the discrepancy between our measurements and the SVRS-derived \ch{HC3N} values may be the method used to determine the disk-averaged profiles from the SVRS dataset. Our disk averaging calculations do not account for the increasing path length through Titan's atmosphere as the distance from the sub-observer point increases.  Accounting for this effect would lead to an increased weighting of the higher latitudes, including the high winter polar latitudes where \ch{HC3N} is enriched.  This effect is possibly most noticeable in \ch{HC3N} due to the strength of the polar enrichment of this molecule.  However, the uncertainties in our \ch{HC3N} measurement are on the order of the measurements themselves, indicating that our retrieval is not very sensitive to this molecule.

\begin{deluxetable}{llllllllllll}
\tabletypesize{\tiny}
\tablecaption{Retrieved volume mixing ratios for major gases at our observations' height of greatest sensitivity, determined by NEMESIS contribution functions, alongside retrieved surface temperature $T_{s}$. Errors represent fitting uncertainties associated with the NEMESIS retrieval scheme. \label{tab:vmrs}}

\tablehead{
\colhead{Observation Date} & \colhead{$T_{s}$ (K)} & \colhead{\ch{CH3D}} & \colhead{\ch{C2H2}} & \colhead{\ch{C2H4}} & \colhead{\ch{C2H6}} & \colhead{\ch{C3H4}}  & 
\colhead{\ch{C3H8}}  & \colhead{\ch{C4H2}} & 
\colhead{\ch{CO2}} &  \colhead{\ch{HCN}} &
\colhead{\ch{HC3N}}\\
\colhead{Height (km)} & \colhead{} & \colhead{154}  & \colhead{124} & \colhead{154} & \colhead{154} & \colhead{154} & \colhead{124} & \colhead{202} & \colhead{124} & \colhead{105} & \colhead{332} \\
\colhead{Scaling Factor} & 
\colhead{}  &
\colhead{$10^{-6}$}  & \colhead{$10^{-6}$} & \colhead{$10^{-7}$} & \colhead{$10^{-5}$} & \colhead{$10^{-8}$} & \colhead{$10^{-7}$} & \colhead{$10^{-9}$} & \colhead{$10^{-8}$} & \colhead{$10^{-7}$} & \colhead{$10^{-8}$} }

\startdata
Initial Values & $93.0\pm1.0$ & $6.60\pm0.66$ & $2.62\pm0.52$ & $1.20\pm0.24$ & $1.27\pm0.13$ & $0.80\pm0.16$ & $3.66\pm0.73$ & $3.03\pm0.61$ & $1.60\pm0.32$ & $1.34\pm0.27$ & $0.21\pm2.11$ \\ \hline
2004-03-05 & $90.1\pm0.3$ & $7.22\pm0.72$ & $2.48\pm0.06$ & $1.82\pm0.08$ & $1.10\pm0.07$ & $0.99\pm0.04$ & $3.66\pm0.20$ & $4.50\pm0.34$ & $2.11\pm0.17$ & $1.17\pm0.07$ & $1.52\pm1.20$ \\
2004-03-25 & $90.9\pm0.3$ & $7.27\pm0.73$ & $2.55\pm0.06$ & $1.81\pm0.08$ & $1.12\pm0.08$ & $1.00\pm0.04$ & $3.52\pm0.19$ & $4.42\pm0.34$ & $2.04\pm0.16$ & $1.16\pm0.07$ & $1.39\pm1.10$ \\
2005-11-15 & $93.0\pm0.3$  & $6.63\pm0.60$ & $2.25\pm0.05$ & $1.62\pm0.06$ & $1.02\pm0.07$ & $0.87\pm0.03$ & $3.38\pm0.18$ & $4.31\pm0.33$ & $1.65\pm0.11$ & $1.02\pm0.05$ & $1.68\pm1.45$ \\
2006-04-20 & $90.7\pm0.3$ & $6.98\pm0.67$ & $2.54\pm0.06$ & $1.66\pm0.07$ & $1.11\pm0.07$ & $0.92\pm0.03$ & $3.49\pm0.19$ & $4.21\pm0.30$ & $2.16\pm0.18$ & $1.10\pm0.06$ & $1.89\pm1.64$ \\
2004-03-04 & $91.1\pm0.3$ & & $2.67\pm0.05$ & $1.73\pm0.05$ & $1.06\pm0.07$ & $1.07\pm0.03$ & $3.24\pm0.14$ & $4.62\pm0.30$ & $2.11\pm0.14$ & $1.20\pm0.06$ & $1.25\pm0.78$ \\
2004-11-16 & $90.6\pm0.3$ & & $2.57\pm0.05$ & $1.67\pm0.06$ & $1.01\pm0.07$ & $1.02\pm0.03$ & $3.18\pm0.14$ & $4.71\pm0.31$ & $1.94\pm0.12$ & $1.12\pm0.05$ & $1.79\pm1.25$ \\
2005-03-23 & $90.8\pm0.3$ & & $2.53\pm0.05$ & $1.63\pm0.05$ & $1.01\pm0.06$ & $0.97\pm0.03$ & $3.11\pm0.13$ & $4.46\pm0.28$ & $2.01\pm0.13$ & $1.06\pm0.05$ & $1.68\pm1.13$ \\
2005-04-16 & $90.6\pm0.3$ & & $2.45\pm0.04$ & $1.65\pm0.05$ & $1.00\pm0.07$ & $0.93\pm0.03$ &  $3.11\pm0.13$ & $4.36\pm0.27$ & $2.00\pm0.13$ & $1.04\pm0.04$ & $1.69\pm1.12$ \\ \hline
SVRS, 2006-01 & & & $1.74\pm0.06$ & $1.84\pm0.34$ & $0.85\pm0.10$ & $0.90\pm0.13$ & & $3.99\pm0.64$ & & $0.94\pm0.38$ & $0.19\pm0.06$
\enddata
\end{deluxetable}

\section{Discussion}
\label{sec:discussion}

\subsection{Comparison to Cassini CIRS Data}
\label{sec:comparison}

Temperature profile retrievals for the Map Dataset were compared to CIRS profiles retrieved in \citet{Teanby2019}.  They retrieved full temperature profiles for Titan's atmosphere at various latitudes and gas VMRs for \ch{C4H2}, \ch{C3H4}, \ch{HC3N}, \ch{CO2}, HCN, \ch{C2H2}, \ch{C2H6}, and \ch{C2H4} at a pressure of 1 mbar (this pressure roughly translated to a 185 km height in our global average) from Jul 2004 to Sept 2017.  Specific dates covered were generally within 4 months of our IRS observations. After finding the \citet{Teanby2019} dataset closest in time to each of our observations, we used data where the observed \citet{Teanby2019} latitude was closest to the sub-observer latitude at the time of observation. Only dates with data points within $10 \degree$ of the sub-observer latitude were used.  This was done to approximate a global average as the \citet{Teanby2019} retrievals were only performed at specific latitudes and generally did not include full hemispheric coverage. Due to only having one data point (at a pressure of 1 mbar) per latitude, VMR data from \citet{Teanby2019} was too sparse to compare in detail to our gas profile retrievals.

\begin{deluxetable*}{llll}
\tabletypesize{\small}
\tablecaption{Summary of the Map Dataset observation dates and sub-observer latitude, $L_{obs}$, alongside the closest date and latitude in \citet{Teanby2019}.\label{tab:teanby2019}}
\tablewidth{0pt}
\tablehead{
\colhead{Spitzer Date} & \colhead{Teanby Date} & \colhead{$L_{obs}$ (\degree)} & \colhead{Teanby $L$ (\degree)}
}
\startdata
2004-03-05 & 2004-07-02 & -26.24 & -25.20 \\
2004-03-25 & 2004-07-02 & -26.28 & -25.20 \\
2005-11-15 & 2005-10-29 & -17.07 & -15.20 \\
2006-04-20 & 2006-05-22 & -19.98 & -20.40 \\
\enddata
\end{deluxetable*}

\begin{figure}
\centering

  \includegraphics[width=0.7\linewidth]{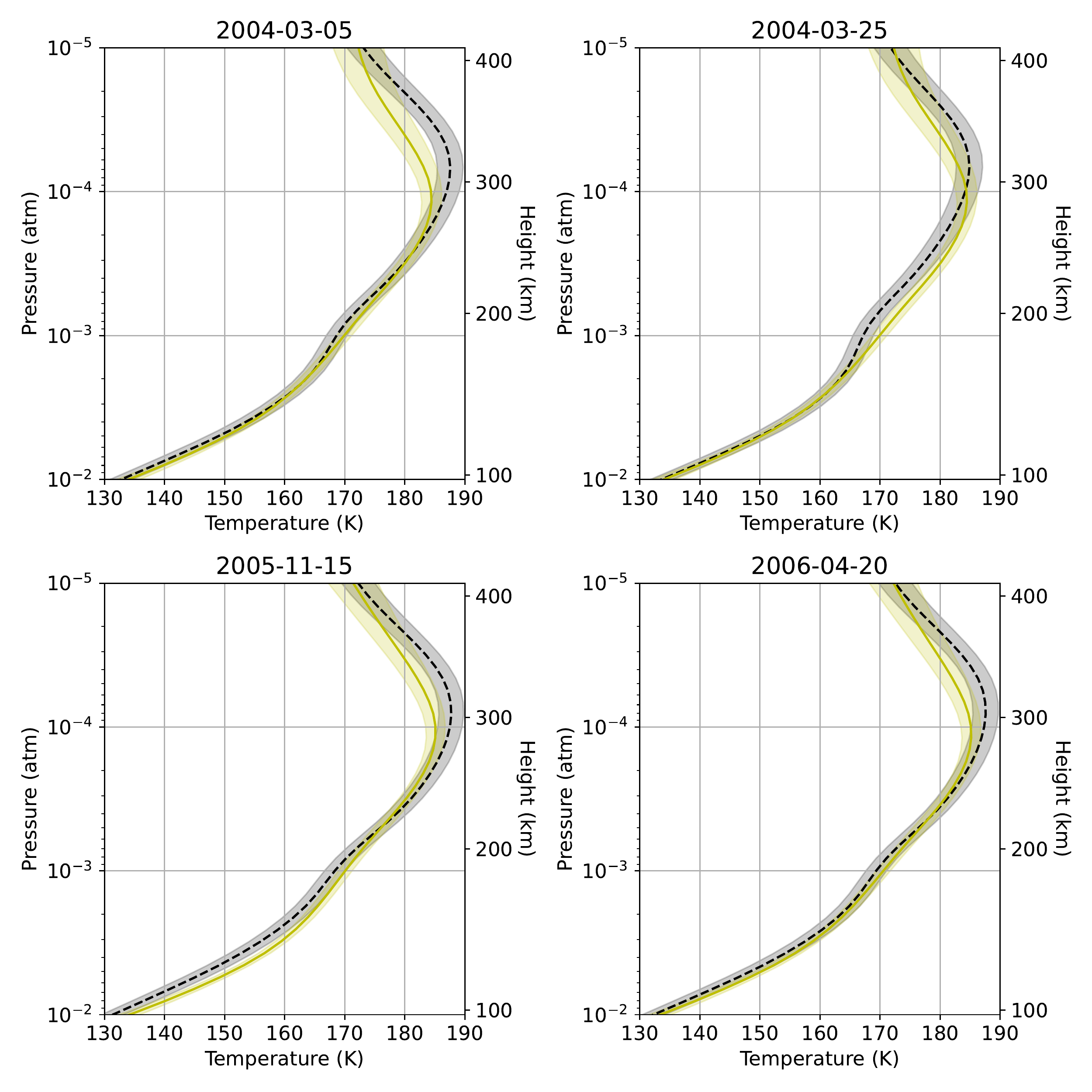}
  
\caption{Disk-averaged temperature retrievals for Spitzer data resulting from NEMESIS spectral fitting (black dashed line) compared to similar Cassini CIRS retrievals (yellow line) at the sub-observer latitudes listed in Table \ref{tab:teanby2019}. Filled regions indicate 1-$\sigma$ CIRS retrieval error (yellow) and IRS retrieval error (black). 
}\label{fig:cirs_temp}
\end{figure}

\label{sec:cirs_temp}
Comparison of our retrieved temperature to that from \citet{Teanby2019} can be seen in Figure \ref{fig:cirs_temp}.  Results are generally in agreement, with overlapping 1-sigma error bars, especially in the high contribution function $100-250$ km height range. The warmer temperatures seen at $300-400$ km from our retrievals may be due to the warm, elevated stratopause poleward of $50\degree$ in the winter hemisphere \citep[e.g.][]{TEANBY2008,achterberg2011} contributing to the overall emission in the Spitzer disk-average spectra, compared to low-latitude temperature profiles from \citet{Teanby2019} which do not have a warm, elevated stratopause.

\subsection{Unmodeled Emission Features and Possible Origins}
\label{sec:features}
\begin{figure}
    \centering
    \includegraphics[width=\linewidth]{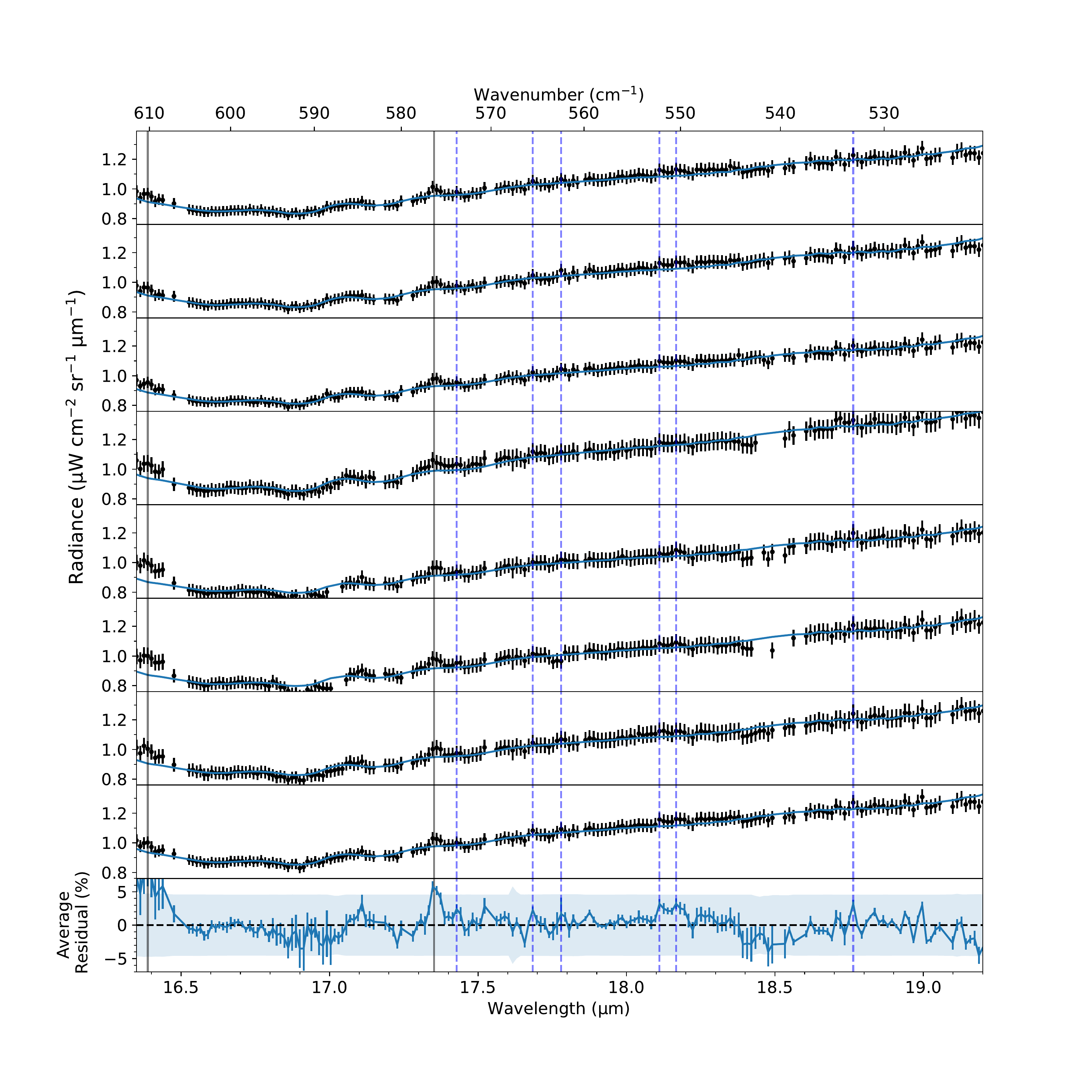}
    \caption{Individual observations (black errorbars) and model fits (blue) alongside averaged residuals (bottom) over the $16.2-\SI{18.2}{\micro\metre}$ range. Errorbars for the average residuals represent the standard deviation of residuals across the eight observations---smaller errorbars represent more consistent features. The shaded blue region represents our estimated noise level. While many narrow spectral features appear to be present across all spectra, the 16.39 and \SI{17.35}{\micro\metre} features (solid lines) are the only candidate features that are consistently above the estimated noise level ($\sim4\%$). Dashed lines at 17.43, 17.68, 17.78, 18.11, 18.17, and \SI{18.76}{\micro\metre} represent possible candidate features that may be better resolved through follow-up observations with JWST.}
    \label{fig:unmodeled}
\end{figure}

We noticed two narrow emission features above our assumed noise level that are not well explained by our model spectrum, centered at \SI{17.35}{\um} (\SI{576.2}{\wn}) and \SI{16.39}{\micro\metre} (\SI{610.2}{\wn}), shown in Figure \ref{fig:unmodeled}.
The \SI{17.35}{\micro\metre} feature was also detected in an IRIS-based analysis \citep{COURTIN1995}.  They postulated that the unidentified emission could be due to a weak \ch{H2O} transition near 576 cm$^{-1}$, but noted that this is unlikely due to the overall low concentration of water vapor in Titan's atmosphere and the fact that stronger \ch{H2O} bands are not present.  Similarly, varying water vapor profiles in our model did not lead to any improvement in fitting the feature. 

\subsubsection{Buckminsterfullerene}
Neutral buckminsterfullerene (\ch{C60}) has a well-constrained emission line that is centered at \SI{17.35}{\micro\metre} when in solid phase \citep{Brieva2016}.  Production of neutral \ch{C60} in Titan's atmosphere has been theorized \citep{SITTLER2020}, but not detected.  \ch{C60} has also been detected in young planetary nebula via IRS observations \citep{Cami2010}. However, it is unlikely that a heavy molecule such as solid \ch{C60} could remain trapped in Titan's atmosphere for a significant amount of time and we were unable to detect the stronger feature associated with \ch{C60} at \SI{18.98}{\micro\metre}. Detection of the stronger feature in future studies would greatly help in the identification of \ch{C60} in the atmosphere.

\subsubsection{Polycyclic Aromatic Hydrocarbons}
Various species of polycyclic aromatic hydrocarbons (PAHs) have been detected in large abundances in Titan's upper atmosphere \citep{Lopez2013}. In addition, certain types of PAHs detected in the interstellar medium exhibit strong spectral features at 16.43 and \SI{17.38}{\micro\metre} with FWHM values of 0.16 and \SI{0.13}{\micro\metre}, respectively \citep{Moutou2000,peeters2004polycyclic}.  However, these two features are redder than the features observed in our data by about three wavelength steps.  The \SI{16.43}{\micro\metre} feature is also predicted to be significantly stronger than the \SI{17.38}{\micro\metre} feature, something not evident in our data.

The \href{https://www.astrochemistry.org/pahdb/}{NASA Ames Research Center PAH IR Spectral Database} \citep{Boersma_2014,Bauschlicher_2018,Mattioda_2020} contains detailed data on prominent IR transitions of over 4000 types of PAHs. It is unlikely that a single neutral PAH is a perfect match for our observed features.  For example, the \ch{C22} molecule's strongest transition is centered at \SI{17.35}{\micro\metre}, but lacks any strong features near \SI{16.39}{\micro\metre}. \ch{C33} contains a strong feature centered at \SI{16.38}{\micro\metre}, but also contains a stronger feature centered at \SI{18.18}{\micro\metre} not consistent with our data.  In addition, the charge of each species shifts where their spectral features are located \citep{bauschlicher2008}. 

\citet{Lopez2013} used the PAH Database to find that a \SI{3.3}{\micro\metre} emission feature in the upper ($900-1000$ km) atmosphere could mostly be explained by high concentrations ($\sim10^{4}/$cm$^{3}$) of the neutral PAHs \ch{C48H22} and \ch{C10H8N}. However, these PAHs do not show significant spectral contribution near our unidentified features. A future principal component analysis (PCA) of the higher signal to noise (S/N) and spectral resolution MIRI data using the Ames PAH Database similar to that presented in \citet{Lopez2013} may be able to identify the combination of individual molecules that are the source of our spectral features.   

Recent studies have attempted to recreate the IR spectral contribution of Titan's organic haze through laboratory experimentation with $\ch{N2}-\ch{CH4}$ gaseous mixtures \citep{DUBOIS2019,perrin2021}, but lack coverage in our $16.0-\SI{19.5}{\micro\metre}$ area of interest. Better understanding of the impact of PAHs to Titan's infrared spectrum through laboratory experimentation or \textit{in situ} exploration through the upcoming \textit{Dragonfly} mission \citep{lorenz2018,barnes2021} will help constrain origin of the unmodeled spectral features we see here.

\subsubsection{Propylene}
The pseudo line list for \ch{C3H6} used in our model \citep{SUNG2018} only includes information in the $6.5-\SI{15.4}{\micro\metre}$ ($600-\SI{1534}{\wn}$) wavelength range. More recent high-resolution cross-section measurements of \ch{C3H6} at longer wavelengths have revealed a spectral band centered at 576 cm$^{-1}$ \citep[\SI{17.36}{\um},][]{bernath2023}. Since this gas has already been detected on Titan in the infrared, this band could potentially explain the missing \SI{17.35}{\micro\metre}  emission. However, to check whether it is a good fit, we require a new low-temperature line list or pseudo line list for this region.

\subsection{JWST Cycle 1}
JWST is scheduled to observe Titan with MIRI \citep{rieke2015mid} in the $5.0-\SI{28.3}{\micro\metre}$ ($350-\SI{2000}{\wn}$) range during its Cycle-1 GTO observations \citep{Nixon2016,nixon2021}. The primary focus of these observations is the $5-\SI{7}{\micro\metre}$ wavelength range, which has not been observed by CIRS nor most other space observatories. 

\begin{figure}
\centering
\begin{subfigure}{.48\textwidth}
  \centering
  \includegraphics[width=\linewidth]{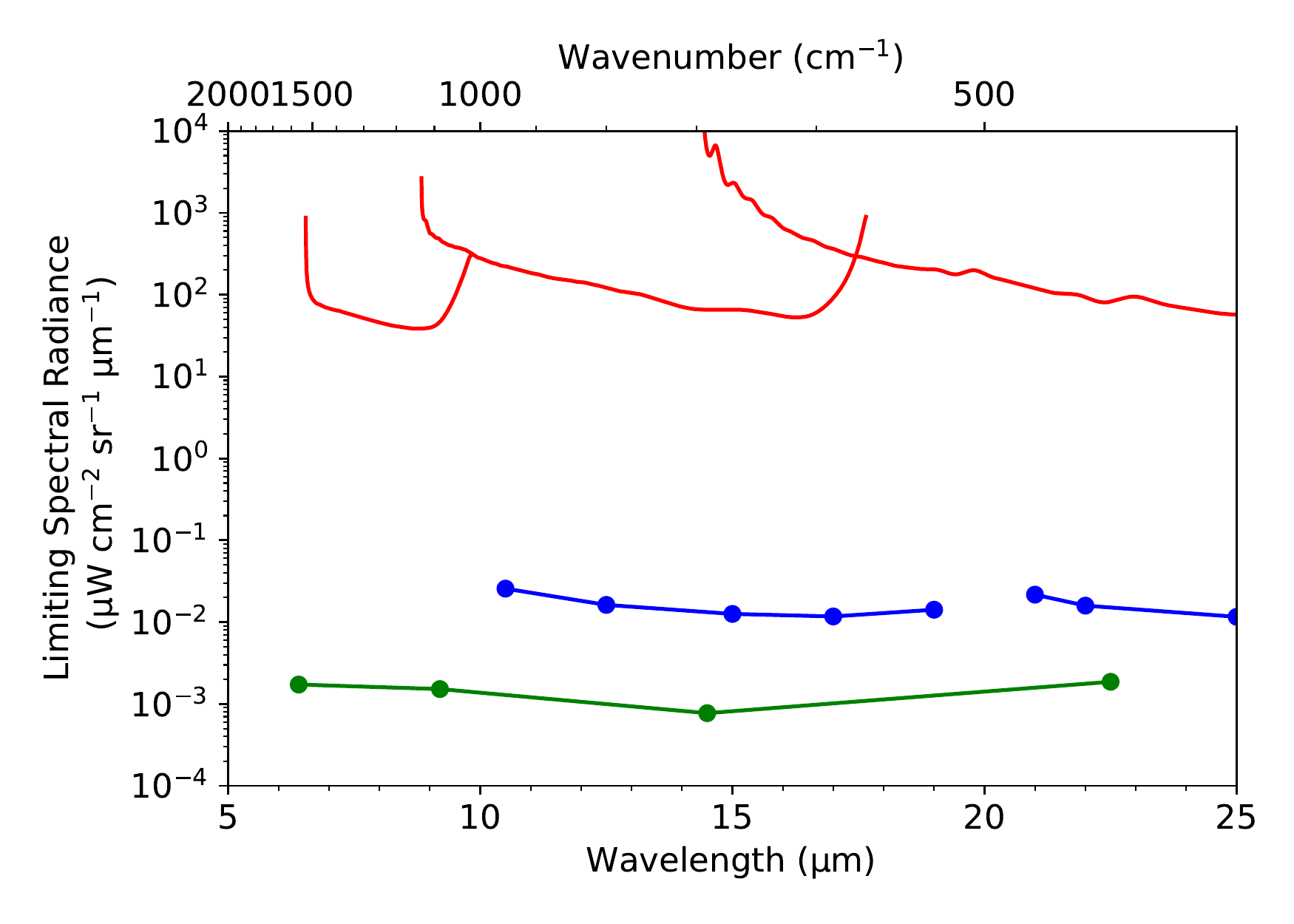}
  \caption{Limiting Signal}
\end{subfigure}
\begin{subfigure}{.48\textwidth}
  \centering
  \includegraphics[width=\linewidth]{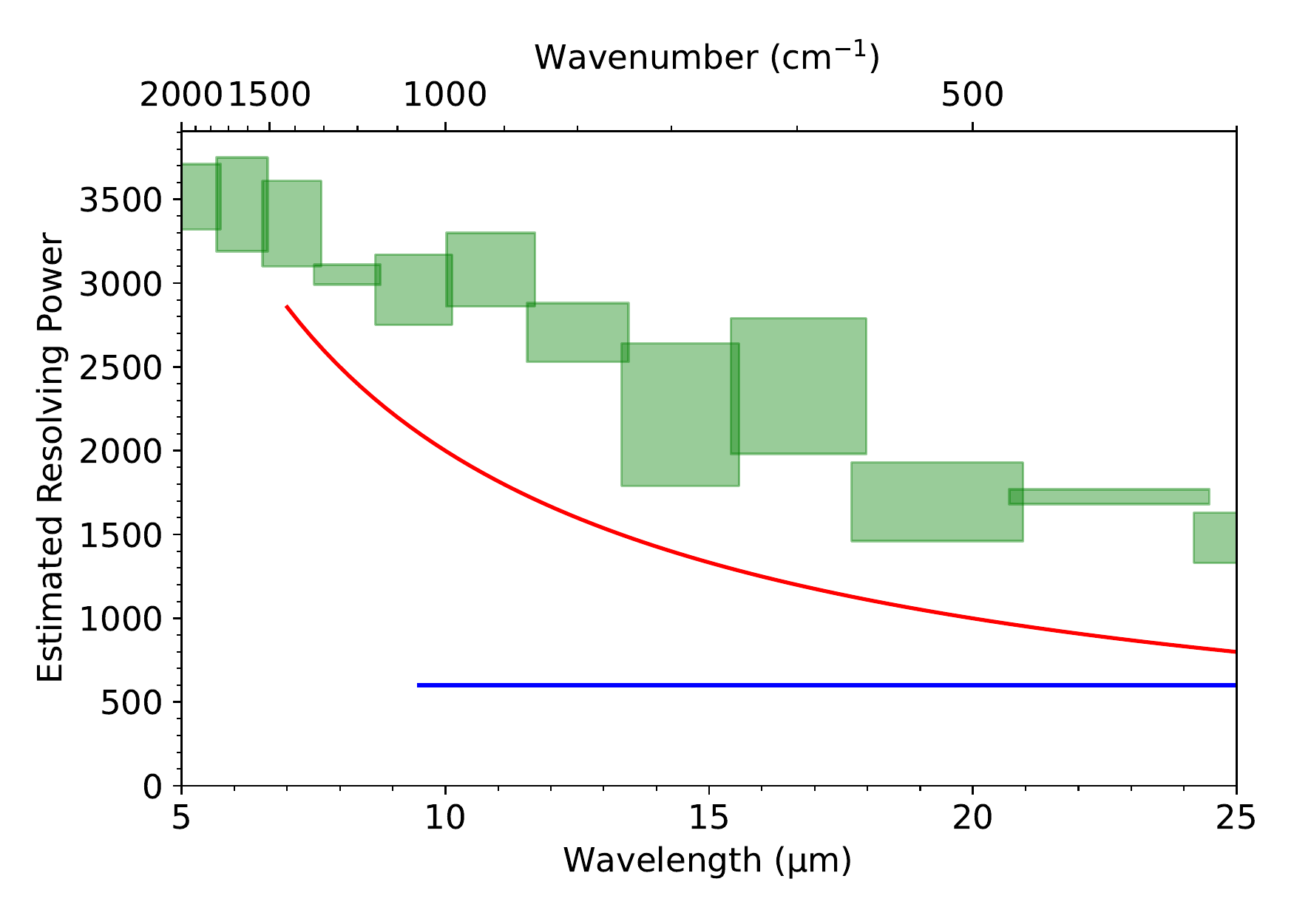}
  \caption{Resolving Power}
\end{subfigure}
\caption{(a) Estimated limiting continuum fluxes for the JWST MIRI (green), Spitzer IRS (blue), and Cassini CIRS (red) high-resolution channels for an S/N$=25$ signal (equivalent to 4\% noise) from a 6 second exposure, based on data compiled by Jane Rigby on the \href{https://www.stsci.edu/jwst/about-jwst/history/historical-sensitivity-estimates}{JWST website} and \protect\citet{flasar2004}. CIRS fluxes were calculated assuming the highest-resolution mode ($\Delta\lambda=\SI{0.5}{\wn}$). Line fluxes (\si{\watt~\metre^{-2}}) were converted to radiance units by assuming an Titanian angular diameter of $\delta_{Titan}=0.81''$. The theoretical limiting flux of IRS ($\gtrsim\SI{1e-2}{\micro W~cm^{-2}~sr^{-1}~\micro m ^{-1}}$) is slightly lower than our assumed error ($\gtrsim\SI{3e-2}{\micro W~cm^{-2}~sr^{-1}~\micro m ^{-1}}$) due to accounting for various other sources of error other than purely randomized detector noise, including wavelength/resolution shifts and uncertainty in haze parameters. (b) Approximate resolving power $R$ for each instrument over the same wavelength range.  MIRI estimates are based on \protect\citet{labiano2021}.}
\label{fig:limiting}
\end{figure}

For a given S/N, MIRI is expected to have flux sensitivities $\sim20-70$ times lower than the Spitzer IRS SH module, with an average of $\sim21$ times higher sensitivity in our $16.35-\SI{19.35}{\micro\metre}$ region of interest (see Figure \ref{fig:limiting}). Many of the spectral features observed in the $16.35-\SI{19.35}{\micro\metre}$ range across our IRS observations are very narrow (FWHM of $\sim2-4$ wavelength steps) and unresolved. MIRI will have a resolving power $\sim2.5-4.5\times$ that of IRS in the region and $\sim1.3-2.5\times$ that of CIRS. Alongside increased resolving power, increased sensitivity will allow for better resolution of the candidate emission features seen in Figure \ref{fig:unmodeled} and help confirm these as real features, allowing for further study of their origins.  This may also help in the identification of many exotic species active in the infrared that are theorized to be produced in the upper atmosphere, as well as aid in further study of previously detected species like \ch{C3H6} while providing better constraints on spatial and temporal variations in their abundances. Better spatial resolution will also help in the study of the time-dependence of the polar enrichment effect described in Section \ref{sec:shfitting}.

\section{Conclusions and Future Work}
\label{sec:conclusion}
We performed the first reduction of both high-resolution and low-resolution IR spectra of Titan taken by the Spitzer Space Telescope and made alterations to the spectral inversion algorithm NEMESIS to retrieve various profiles for the disk-averaged spectra of Titan.  The data reduction process used in this work largely builds upon \citet{Rowe_Gurney_2021} and \citet{ORTON2014a,ORTON2014b}. Spitzer data largely confirmed results derived from previous observations from ISO and CIRS, as well as haze properties derived in \citet{Tomasko2008}, \citet{ANDERSON2011}, and \citet{VINATIER2012}.  Limitations due to the spatial resolution of Spitzer when compared to that of previous observations made it impossible to measure spatial variations in our gas and temperature profiles. 

Issues still present in fitting well-known spectral bands, especially in the methane, ethane, and acetylene band peaks suggest that more work is needed regarding calibration of IRS observations.  Testing of different fitting techniques for the resolving power used in the telescope filter functions required for $k$-table creation could lead to more accurate results. Better constraints on the true error of IRS data on bright sources would also help us identify more regions of the spectra with fitting issues.

We observed unidentified strong emission features at \SI{16.39}{\micro\metre} and \SI{17.35}{\micro\metre} and postulated possible chemical species that could explain these discrepancies, including \ch{H2O}, \ch{C60}, \ch{C3H6}, and various PAH species.  Several other narrow emission features in the $16.00-\SI{19.35}{\micro\metre}$ range were observed across multiple spectra but cannot be confirmed given the moderate resolving power and S/N of IRS.  Future investigation of these features through JWST observations used in tandem with pseudo-line lists for the above chemical species will help in narrowing down their origins as well as improve our understanding of the effects of Titan's haze on our retrieved spectra.  

The results we present here lay the ground work for JWST MIRI observations of Titan in a spectral window that has been underutilized due to poor data sensitivity in the existing dataset.  We propose the following steps be taken to maximize the scientific products from future observations of these spectral windows:
\begin{enumerate}
    \item The spectral region from 16 to \SI{20}{\  um} lacks low temperature laboratory spectra of molecules present in Titan's atmosphere.  Cross-sectional measurements should be made to enable the qualitative detection of these molecules in this spectral window from existing Spitzer IRS and future JWST MIRI observations.
    \item Pseudo line lists, similar to those presented in \citet{Sung2013} and \citet{SUNG2018}, for detected and predicted species would enable the proper fitting and retrieval of molecular abundance profiles from these spectral windows.  Additionally, these emission bands may be sensitive to different altitude ranges than those observed at higher wavenumbers.
    \item JWST MIRI observations of these spectral windows, with the appropriate updated laboratory cross sections and pseudo line lists, would potentially enable a more complete vertical description of molecular abundance from ground and space-based observatories \citep[e.g.][]{Lombardo2019b}.
    \item Observations from these spectral windows would help to constrain the hemispheric asymmetry of Titan's stratospheric composition due to the polar enrichment caused by the stratospheric meridional overturning circulation.  These constraints would help to elucidate the dynamical changes occurring in Titan's atmosphere as it progresses through its seasonal changes.
\end{enumerate}

\section*{Acknowledgements}

\begin{acknowledgments}

B.P.C. and C.A.N. were funded by the NASA Astrobiology Institute.  B.P.C., N.R.G., and R.A. were further supported by the Center for Research and Exploration in Space Science \& Technology II (CRESST II) under NASA award number 80GSFC21M0002.  N.R.G. and L.N.F. were supported by a European Research Council Consolidator Grant (under the EU’s Horizon 2020 research and innovation program, grant agreement No. 723890) at the University of Leicester.  This work is based on observations made with the Spitzer Space Telescope, which is operated by the Jet Propulsion Laboratory, California Institute of Technology under a contract with NASA. This research has made use of the NASA/IPAC Infrared Science Archive \citep{https://doi.org/10.26131/irsa543}, which is funded by the National Aeronautics and Space Administration and operated by the California Institute of Technology. Extinction cross-section measurements for \ch{C3H6} were compiled by Peter Bernath at the Department of Chemistry and Biochemistry at Old Dominion University. The authors give special thanks to Jan Cami of the University of Western Ontario, who was consulted on the spectral features of fullerenes and PAHs and the viability of their existence in Titan’s atmosphere.

\end{acknowledgments}

\section*{Data Availability}
All data used in this analysis are publicly available online at the \href{https://sha.ipac.caltech.edu/applications/Spitzer/SHA/}{Spitzer Heritage Archive} \citep{https://doi.org/10.26131/irsa430} by searching  ``Titan" in the ``Moving Object'' search option. 

\appendix

\section{LH Data}
\begin{figure}
    
    \centering
    \includegraphics[width=\linewidth]{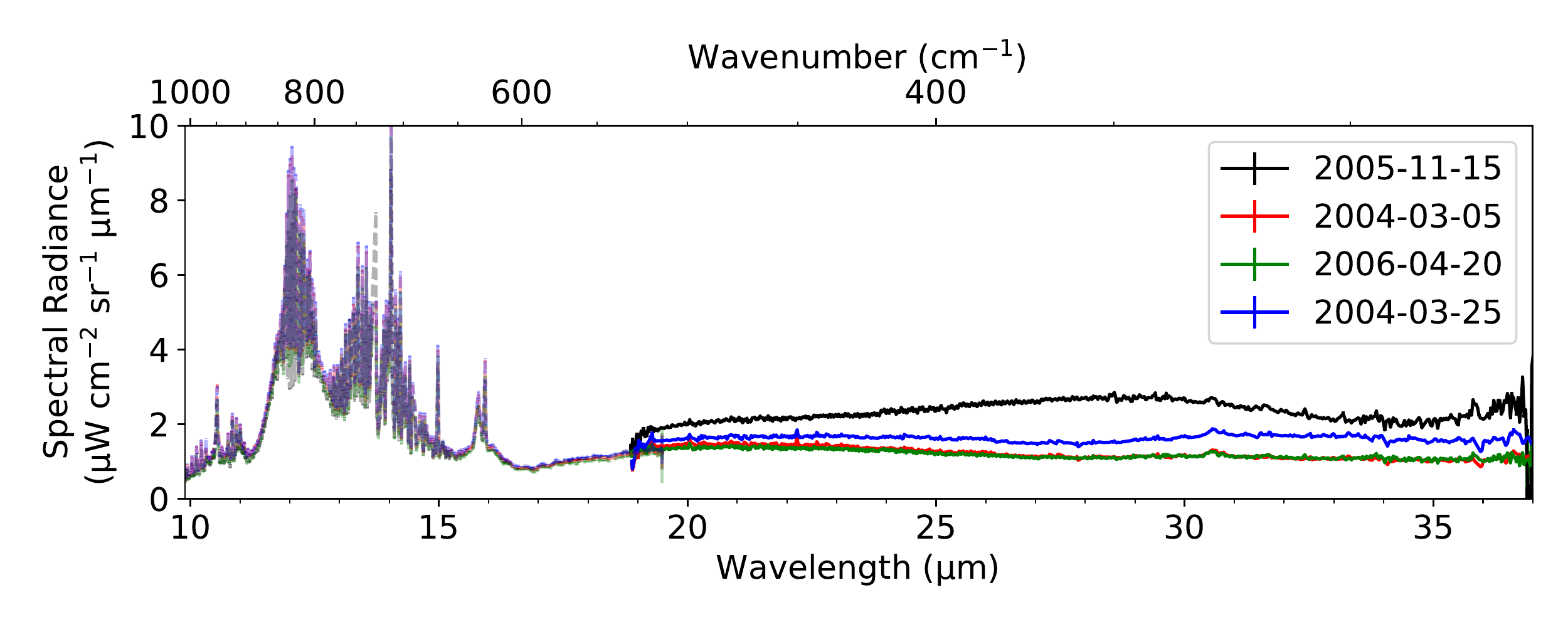}
    \caption{Automated spectra output from the IRS pipeline for the SH (dashed) module alongside their LH (solid) counterparts for Map Dataset 1. Only observations from 2004/03/05 and 2006/04/20 are generally consistent with SL data and each other and may be usable for future analyses of LH data.}
    \label{fig:lh}
\end{figure}

Automated spectra produced by the IRS pipeline (and converted from flux density units to radiance units) for LH data are shown in Figure \ref{fig:lh}.  The LH data were largely inconsistent across observations and discontinuous with SL in overlapping regions---noise reduction techniques were ineffective in changing this result.  The reason for these vast discrepancies is not clear. MIRI will allow for better study of the $19.5-\SI{28.0}{\micro\metre}$ covered solely by the LH module.

\bibliography{ref}

\begin{thebibliography}{}
\expandafter\ifx\csname natexlab\endcsname\relax\def\natexlab#1{#1}\fi
\providecommand{\url}[1]{\href{#1}{#1}}
\providecommand{\dodoi}[1]{doi:~\href{http://doi.org/#1}{\nolinkurl{#1}}}
\providecommand{\doeprint}[1]{\href{http://ascl.net/#1}{\nolinkurl{http://ascl.net/#1}}}
\providecommand{\doarXiv}[1]{\href{https://arxiv.org/abs/#1}{\nolinkurl{https://arxiv.org/abs/#1}}}

\bibitem[{Achterberg {et~al.}(2011)Achterberg, Gierasch, Conrath, Flasar, \&
  Nixon}]{achterberg2011}
Achterberg, R.~K., Gierasch, P.~J., Conrath, B.~J., Flasar, F.~M., \& Nixon,
  C.~A. 2011, Icarus, 211, 686

\bibitem[{Anderson \& Samuelson(2011)}]{ANDERSON2011}
Anderson, C.~M., \& Samuelson, R.~E. 2011, Icarus, 212, 762,
  \dodoi{https://doi.org/10.1016/j.icarus.2011.01.024}

\bibitem[{Barnes {et~al.}(2021)Barnes, Turtle, Trainer, Lorenz, MacKenzie,
  Brinckerhoff, Cable, Ernst, Freissinet, Hand, {et~al.}}]{barnes2021}
Barnes, J.~W., Turtle, E.~P., Trainer, M.~G., {et~al.} 2021, The Planetary
  Science Journal, 2, 130

\bibitem[{Bauschlicher {et~al.}(2018)Bauschlicher, Ricca, Boersma, \&
  Allamandola}]{Bauschlicher_2018}
Bauschlicher, C.~W., Ricca, A., Boersma, C., \& Allamandola, L.~J. 2018, The
  Astrophysical Journal Supplement Series, 234, 32,
  \dodoi{10.3847/1538-4365/aaa019}

\bibitem[{Bauschlicher~Jr {et~al.}(2008)Bauschlicher~Jr, Peeters, \&
  Allamandola}]{bauschlicher2008}
Bauschlicher~Jr, C.~W., Peeters, E., \& Allamandola, L.~J. 2008, The
  Astrophysical Journal, 678, 316

\bibitem[{Bernath {et~al.}(2023)Bernath, Dodangodage, Zhao, \&
  Billinghurst}]{bernath2023}
Bernath, P.~F., Dodangodage, R., Zhao, J., \& Billinghurst, B. 2023, Journal of
  Quantitative Spectroscopy and Radiative Transfer, 296, 108462

\bibitem[{Boersma {et~al.}(2014)Boersma, Bauschlicher, Ricca, Mattioda, Cami,
  Peeters, de~Armas, Saborido, Hudgins, \& Allamandola}]{Boersma_2014}
Boersma, C., Bauschlicher, C.~W., Ricca, A., {et~al.} 2014, The Astrophysical
  Journal Supplement Series, 211, 8, \dodoi{10.1088/0067-0049/211/1/8}

\bibitem[{Brieva {et~al.}(2016)Brieva, Gredel, Jäger, Huisken, \&
  Henning}]{Brieva2016}
Brieva, A.~C., Gredel, R., Jäger, C., Huisken, F., \& Henning, T. 2016, The
  Astrophysical Journal, 826, 122, \dodoi{10.3847/0004-637x/826/2/122}

\bibitem[{Burgdorf {et~al.}(2006)Burgdorf, Orton, {van Cleve}, Meadows, \&
  Houck}]{BURGDORF2006}
Burgdorf, M., Orton, G., {van Cleve}, J., Meadows, V., \& Houck, J. 2006,
  Icarus, 184, 634, \dodoi{https://doi.org/10.1016/j.icarus.2006.06.006}

\bibitem[{Cami {et~al.}(2010)Cami, Bernard-Salas, Peeters, \& Malek}]{Cami2010}
Cami, J., Bernard-Salas, J., Peeters, E., \& Malek, S. 2010, Science, 329,
  1180, \dodoi{10.1126/science.1192035}

\bibitem[{Chedin {et~al.}(1982)Chedin, Husson, \& Scott}]{chedin}
Chedin, A., Husson, N., \& Scott, N. 1982, Bulletin d'Information du Centre de
  Donnees Stellaires, 22, 121

\bibitem[{Cottini {et~al.}(2012)Cottini, Nixon, Jennings, Anderson, Gorius,
  Bjoraker, Coustenis, Teanby, Achterberg, Bézard, {de Kok}, Lellouch, Irwin,
  Flasar, \& Bampasidis}]{COTTINI2012}
Cottini, V., Nixon, C., Jennings, D., {et~al.} 2012, Icarus, 220, 855,
  \dodoi{https://doi.org/10.1016/j.icarus.2012.06.014}

\bibitem[{Courtin {et~al.}(1995)Courtin, Gautier, \& McKay}]{COURTIN1995}
Courtin, R., Gautier, D., \& McKay, C.~P. 1995, Icarus, 114, 144,
  \dodoi{https://doi.org/10.1006/icar.1995.1050}

\bibitem[{{Coustenis} {et~al.}(2003){Coustenis}, {Salama}, {Schulz}, {Ott},
  {Lellouch}, {Encrenaz}, {Gautier}, \& {Feuchtgruber}}]{Coustenis2003}
{Coustenis}, A., {Salama}, A., {Schulz}, B., {et~al.} 2003, \icarus, 161, 383,
  \dodoi{10.1016/S0019-1035(02)00028-3}

\bibitem[{{Coustenis} {et~al.}(1998){Coustenis}, {Salama}, {Lellouch},
  {Encrenaz}, {Bjoraker}, {Samuelson}, {de Graauw}, {Feuchtgruber}, \&
  {Kessler}}]{Coustenis1998}
{Coustenis}, A., {Salama}, A., {Lellouch}, E., {et~al.} 1998, \aap, 336, L85

\bibitem[{Coustenis {et~al.}(2002)Coustenis, Encrenaz, Lellouch, Salama,
  Müller, Burgdorf, Schmitt, Feuchtgruber, Schulz, Ott, {de Graauw}, Griffin,
  \& Kessler}]{COUSTENIS2002}
Coustenis, A., Encrenaz, T., Lellouch, E., {et~al.} 2002, Advances in Space
  Research, 30, 1971, \dodoi{https://doi.org/10.1016/S0273-1177(02)00577-X}

\bibitem[{{Coustenis} {et~al.}(2007){Coustenis}, {Achterberg}, {Conrath},
  {Jennings}, {Marten}, {Gautier}, {Nixon}, {Flasar}, {Teanby}, {B{\'e}zard},
  {Samuelson}, {Carlson}, {Lellouch}, {Bjoraker}, {Romani}, {Taylor}, {Irwin},
  {Fouchet}, {Hubert}, {Orton}, {Kunde}, {Vinatier}, {Mondellini}, {Abbas}, \&
  {Courtin}}]{Coustenis2007}
{Coustenis}, A., {Achterberg}, R.~K., {Conrath}, B.~J., {et~al.} 2007, \icarus,
  189, 35, \dodoi{10.1016/j.icarus.2006.12.022}

\bibitem[{Creecy {et~al.}(2019)Creecy, Li, Jiang, Nixon, West, \&
  Kenyon}]{creecy2019}
Creecy, E.~C., Li, L., Jiang, X., {et~al.} 2019, Geophysical Research Letters,
  46, 13649

\bibitem[{Dobrijevic {et~al.}(2014)Dobrijevic, HÉBRARD, Loison, \&
  Hickson}]{Dobrijevic2014}
Dobrijevic, M., HÉBRARD, E., Loison, J.-C., \& Hickson, K. 2014, Icarus, 228,
  324–346, \dodoi{10.1016/j.icarus.2013.10.015}

\bibitem[{Dubois {et~al.}(2019)Dubois, Carrasco, Petrucciani, Vettier, Tigrine,
  \& Pernot}]{DUBOIS2019}
Dubois, D., Carrasco, N., Petrucciani, M., {et~al.} 2019, Icarus, 317, 182,
  \dodoi{https://doi.org/10.1016/j.icarus.2018.07.006}

\bibitem[{Fazio {et~al.}(2004)Fazio, Hora, Allen, Ashby, Barmby, Deutsch,
  Huang, Kleiner, Marengo, Megeath, {et~al.}}]{fazio2004}
Fazio, G., Hora, J., Allen, L., {et~al.} 2004, The Astrophysical Journal
  Supplement Series, 154, 10

\bibitem[{Flasar {et~al.}(2004)Flasar, Kunde, Abbas, Achterberg, Ade, Barucci,
  B{\'e}zard, Bjoraker, Brasunas, Calcutt, {et~al.}}]{flasar2004}
Flasar, F.~M., Kunde, V., Abbas, M., {et~al.} 2004, The Cassini-Huygens
  Mission, 169

\bibitem[{Giorgini {et~al.}(1996)Giorgini, Yeomans, Chamberlin, Chodas,
  Jacobson, Keesey, Lieske, Ostro, Standish, \& Wimberly}]{Giorgini1996}
Giorgini, J., Yeomans, D., Chamberlin, A., {et~al.} 1996, Bulletin of the
  American Astronomical Society, 28, 1158

\bibitem[{Gordon {et~al.}(2021)Gordon, Rothman, Hargreaves, Hashemi, Karlovets,
  Skinner, Conway, Hill, Kochanov, Tan, Wcisło, Finenko, Nelson, Bernath,
  Birk, Boudon, Campargue, Chance, Coustenis, Drouin, Flaud, Gamache, Hodges,
  Jacquemart, Mlawer, Nikitin, Perevalov, Rotger, Tennyson, Toon, Tran,
  Tyuterev, Adkins, Baker, Barbe, Canè, Császár, Dudaryonok, Egorov,
  Fleisher, Fleurbaey, Foltynowicz, Furtenbacher, Harrison, Hartmann, Horneman,
  Huang, Karman, Karns, Kassi, Kleiner, Kofman, Kwabia–Tchana, Lavrentieva,
  Lee, Long, Lukashevskaya, Lyulin, Makhnev, Matt, Massie, Melosso,
  Mikhailenko, Mondelain, Müller, Naumenko, Perrin, Polyansky, Raddaoui,
  Raston, Reed, Rey, Richard, Tóbiás, Sadiek, Schwenke, Starikova, Sung,
  Tamassia, Tashkun, Auwera, Vasilenko, Vigasin, Villanueva, Vispoel, Wagner,
  Yachmenev, \& Yurchenko}]{GORDON2021}
Gordon, I., Rothman, L., Hargreaves, R., {et~al.} 2021, Journal of Quantitative
  Spectroscopy and Radiative Transfer, 107949,
  \dodoi{https://doi.org/10.1016/j.jqsrt.2021.107949}

\bibitem[{{Houck} \& {van Cleve}(2004)}]{Houck2004b}
{Houck}, J.~R., \& {van Cleve}, J. 2004, {Observations of Outer Solar System
  Satellites and Planets}, Spitzer Proposal

\bibitem[{{Houck} {et~al.}(2004){Houck}, {Roellig}, {van Cleve}, {Forrest},
  {Herter}, {Lawrence}, {Matthews}, {Reitsema}, {Soifer}, {Watson}, {Weedman},
  {Huisjen}, {Troeltzsch}, {Barry}, {Bernard-Salas}, {Blacken}, {Brandl},
  {Charmandaris}, {Devost}, {Gull}, {Hall}, {Henderson}, {Higdon}, {Pirger},
  {Schoenwald}, {Sloan}, {Uchida}, {Appleton}, {Armus}, {Burgdorf},
  {Fajardo-Acosta}, {Grillmair}, {Ingalls}, {Morris}, \&
  {Teplitz}}]{Houck2004a}
{Houck}, J.~R., {Roellig}, T.~L., {van Cleve}, J., {et~al.} 2004, \apjs, 154,
  18, \dodoi{10.1086/423134}

\bibitem[{Hourdin {et~al.}(2004)Hourdin, Lebonnois, Luz, \&
  Rannou}]{hourdin2004}
Hourdin, F., Lebonnois, S., Luz, D., \& Rannou, P. 2004, Journal of Geophysical
  Research: Planets, 109, \dodoi{https://doi.org/10.1029/2004JE002282}

\bibitem[{Husson {et~al.}(1992)Husson, Bonnet, Scott, \& Chedin}]{HUSSON1992}
Husson, N., Bonnet, B., Scott, N., \& Chedin, A. 1992, Journal of Quantitative
  Spectroscopy and Radiative Transfer, 48, 509,
  \dodoi{https://doi.org/10.1016/0022-4073(92)90116-L}

\bibitem[{Hörst {et~al.}(2008)Hörst, Vuitton, \& Yelle}]{horst2008}
Hörst, S.~M., Vuitton, V., \& Yelle, R.~V. 2008, Journal of Geophysical
  Research: Planets, 113, \dodoi{https://doi.org/10.1029/2008JE003135}

\bibitem[{{IRSA}(2022)}]{https://doi.org/10.26131/irsa543}
{IRSA}. 2022, Spitzer Heritage Archive,  IPAC, \dodoi{10.26131/IRSA543}

\bibitem[{{Irwin} {et~al.}(2008){Irwin}, {Teanby}, {de Kok}, {Fletcher},
  {Howett}, {Tsang}, {Wilson}, {Calcutt}, {Nixon}, \& {Parrish}}]{Nemesis}
{Irwin}, P.~G.~J., {Teanby}, N.~A., {de Kok}, R., {et~al.} 2008, \jqsrt, 109,
  1136, \dodoi{10.1016/j.jqsrt.2007.11.006}

\bibitem[{Jacquinet-Husson {et~al.}(2016)Jacquinet-Husson, Armante, Scott,
  Chédin, Crépeau, Boutammine, Bouhdaoui, Crevoisier, Capelle, Boonne,
  Poulet-Crovisier, Barbe, {Chris Benner}, Boudon, Brown, Buldyreva, Campargue,
  Coudert, Devi, Down, Drouin, Fayt, Fittschen, Flaud, Gamache, Harrison, Hill,
  Hodnebrog, Hu, Jacquemart, Jolly, Jiménez, Lavrentieva, Liu, Lodi, Lyulin,
  Massie, Mikhailenko, Müller, Naumenko, Nikitin, Nielsen, Orphal, Perevalov,
  Perrin, Polovtseva, Predoi-Cross, Rotger, Ruth, Yu, Sung, Tashkun, Tennyson,
  Tyuterev, {Vander Auwera}, Voronin, \& Makie}]{JACQUINETHUSSON2016}
Jacquinet-Husson, N., Armante, R., Scott, N., {et~al.} 2016, Journal of
  Molecular Spectroscopy, 327, 31,
  \dodoi{https://doi.org/10.1016/j.jms.2016.06.007}

\bibitem[{Jakobsen {et~al.}(2022)Jakobsen, Ferruit, de~Oliveira, Arribas,
  Bagnasco, Barho, Beck, Birkmann, B{\"o}ker, Bunker, {et~al.}}]{jakobsen2022}
Jakobsen, P., Ferruit, P., de~Oliveira, C.~A., {et~al.} 2022, Astronomy \&
  Astrophysics, 661, A80

\bibitem[{{Jennings} {et~al.}(2009){Jennings}, {Flasar}, {Kunde}, {Samuelson},
  {Pearl}, {Nixon}, {Carlson}, {Mamoutkine}, {Brasunas}, {Guandique},
  {Achterberg}, {Bjoraker}, {Romani}, {Segura}, {Albright}, {Elliott},
  {Tingley}, {Calcutt}, {Coustenis}, \& {Courtin}}]{Jennings2009}
{Jennings}, D.~E., {Flasar}, F.~M., {Kunde}, V.~G., {et~al.} 2009, \apjl, 691,
  L103, \dodoi{10.1088/0004-637X/691/2/L103}

\bibitem[{Jennings {et~al.}(2017)Jennings, Flasar, Kunde, Nixon, Segura,
  Romani, Gorius, Albright, Brasunas, Carlson, Mamoutkine, Guandique,
  Kaelberer, Aslam, Achterberg, Bjoraker, Anderson, Cottini, Pearl, Smith,
  Hesman, Barney, Calcutt, Vellacott, Spilker, Edgington, Brooks, Ade,
  Schinder, Coustenis, Courtin, Michel, Fettig, Pilorz, \&
  Ferrari}]{Jennings:17}
Jennings, D.~E., Flasar, F.~M., Kunde, V.~G., {et~al.} 2017, Appl. Opt., 56,
  5274, \dodoi{10.1364/AO.56.005274}

\bibitem[{{Kunde} {et~al.}(1981){Kunde}, {Aikin}, {Hanel}, {Jennings},
  {Maguire}, \& {Samuelson}}]{Kunde1981}
{Kunde}, V.~G., {Aikin}, A.~C., {Hanel}, R.~A., {et~al.} 1981, \nat, 292, 686,
  \dodoi{10.1038/292686a0}

\bibitem[{{Labiano} {et~al.}(2021){Labiano}, {Argyriou},
  {{\'A}lvarez-M{\'a}rquez}, {Glasse}, {Glauser}, {Patapis}, {Law}, {Brandl},
  {Justtanont}, {Lahuis}, {Mart{\'\i}nez-Galarza}, {Mueller}, {Noriega-Crespo},
  {Royer}, {Shaughnessy}, \& {Vandenbussche}}]{labiano2021}
{Labiano}, A., {Argyriou}, I., {{\'A}lvarez-M{\'a}rquez}, J., {et~al.} 2021,
  \aap, 656, A57, \dodoi{10.1051/0004-6361/202140614}

\bibitem[{{Lacis} \& {Oinas}(1991)}]{Lacis}
{Lacis}, A.~A., \& {Oinas}, V. 1991, \jgr, 96, 9027, \dodoi{10.1029/90JD01945}

\bibitem[{{Lavvas} {et~al.}(2008){Lavvas}, {Coustenis}, \&
  {Vardavas}}]{Lavvas2008}
{Lavvas}, P.~P., {Coustenis}, A., \& {Vardavas}, I.~M. 2008, \planss, 56, 67,
  \dodoi{10.1016/j.pss.2007.05.027}

\bibitem[{Li(2015)}]{li2015}
Li, L. 2015, Scientific Reports, 5, 1

\bibitem[{Li {et~al.}(2011)Li, Nixon, Achterberg, Smith, Gorius, Jiang,
  Conrath, Gierasch, Simon-Miller, Michael~Flasar, Baines, Ingersoll, West,
  Vasavada, \& Ewald}]{Li2011}
Li, L., Nixon, C.~A., Achterberg, R.~K., {et~al.} 2011, Geophysical Research
  Letters, 38, \dodoi{https://doi.org/10.1029/2011GL050053}

\bibitem[{Loison {et~al.}(2019)Loison, Dobrijevic, \& Hickson}]{Loison2019}
Loison, J.-C., Dobrijevic, M., \& Hickson, K. 2019, Icarus, 329,
  \dodoi{10.1016/j.icarus.2019.03.024}

\bibitem[{{Loison} {et~al.}(2015){Loison}, {H{\'e}brard}, {Dobrijevic},
  {Hickson}, {Caralp}, {Hue}, {Gronoff}, {Venot}, \&
  {B{\'e}nilan}}]{Loisin2015}
{Loison}, J.~C., {H{\'e}brard}, E., {Dobrijevic}, M., {et~al.} 2015, \icarus,
  247, 218, \dodoi{10.1016/j.icarus.2014.09.039}

\bibitem[{Lombardo \& Lora(2022)}]{svrs}
Lombardo, N.~A., \& Lora, J.~M. 2022, The Titan Seasonally Varying Radiative
  Species (SVRS) Dataset, 1.0.0,  Zenodo, \dodoi{10.5281/zenodo.7222719}

\bibitem[{Lombardo \& Lora(2023)}]{LOMBARDO2023}
---. 2023, Icarus, 390, 115291,
  \dodoi{https://doi.org/10.1016/j.icarus.2022.115291}

\bibitem[{{Lombardo} {et~al.}(2019{\natexlab{a}}){Lombardo}, {Nixon},
  {Achterberg}, {Jolly}, {Sung}, {Irwin}, \& {Flasar}}]{Lombardo2019a}
{Lombardo}, N.~A., {Nixon}, C.~A., {Achterberg}, R.~K., {et~al.}
  2019{\natexlab{a}}, \icarus, 317, 454, \dodoi{10.1016/j.icarus.2018.08.027}

\bibitem[{{Lombardo} {et~al.}(2019{\natexlab{b}}){Lombardo}, {Nixon},
  {Sylvestre}, {Jennings}, {Teanby}, {Irwin}, \& {Flasar}}]{Lombardo2019b}
{Lombardo}, N.~A., {Nixon}, C.~A., {Sylvestre}, M., {et~al.}
  2019{\natexlab{b}}, \aj, 157, 160, \dodoi{10.3847/1538-3881/ab0e07}

\bibitem[{Lombardo {et~al.}(2019)Lombardo, Nixon, Greathouse, Bézard, Jolly,
  Vinatier, Teanby, Richter, G~Irwin, Coustenis, \& et~al.}]{Lombardo2019c}
Lombardo, N.~A., Nixon, C.~A., Greathouse, T.~K., {et~al.} 2019, The
  Astrophysical Journal, 881, L33, \dodoi{10.3847/2041-8213/ab3860}

\bibitem[{Lopez-Puertas {et~al.}(2013)Lopez-Puertas, Dinelli, Adriani, Funke,
  Garcia-Comas, Moriconi, D'Aversa, Boersma, \& Allamandola}]{Lopez2013}
Lopez-Puertas, M., Dinelli, B., Adriani, A., {et~al.} 2013, The Astrophysical
  Journal, 770, 132, \dodoi{10.1088/0004-637X/770/2/132}

\bibitem[{Lorenz {et~al.}(2018)Lorenz, Turtle, Barnes, Trainer, Adams, Hibbard,
  Sheldon, Zacny, Peplowski, Lawrence, {et~al.}}]{lorenz2018}
Lorenz, R.~D., Turtle, E.~P., Barnes, J.~W., {et~al.} 2018, Johns Hopkins APL
  Technical Digest, 34, 14

\bibitem[{Maguire {et~al.}(1981)Maguire, Hanel, Jennings, Kunde, \&
  Samuelson}]{maguire1981}
Maguire, W., Hanel, R., Jennings, D., Kunde, V., \& Samuelson, R. 1981, Nature,
  292, 683

\bibitem[{Math{\'e} {et~al.}(2020)Math{\'e}, Vinatier, B{\'e}zard, Lebonnois,
  Gorius, Jennings, Mamoutkine, Guandique, \& d’Ollone}]{Mathe2020}
Math{\'e}, C., Vinatier, S., B{\'e}zard, B., {et~al.} 2020, Icarus, 344,
  113547, \dodoi{https://doi.org/10.1016/j.icarus.2019.113547}

\bibitem[{Mattioda {et~al.}(2020)Mattioda, Hudgins, Boersma, Bauschlicher,
  Ricca, Cami, Peeters, de~Armas, Saborido, \& Allamandola}]{Mattioda_2020}
Mattioda, A.~L., Hudgins, D.~M., Boersma, C., {et~al.} 2020, The Astrophysical
  Journal Supplement Series, 251, 22, \dodoi{10.3847/1538-4365/abc2c8}

\bibitem[{McKay {et~al.}(1991)McKay, Pollack, \& Courtin}]{Mckay1991}
McKay, C., Pollack, J., \& Courtin, R. 1991, Science, 253, 1118,
  \dodoi{10.1126/science.253.5024.1118}

\bibitem[{Meadows {et~al.}(2008)Meadows, Orton, Line, Liang, Yung, {Van Cleve},
  \& Burgdorf}]{Meadows2008}
Meadows, V.~S., Orton, G., Line, M., {et~al.} 2008, Icarus, 197, 585,
  \dodoi{https://doi.org/10.1016/j.icarus.2008.05.023}

\bibitem[{{Moutou} {et~al.}(2000){Moutou}, {Verstraete}, {L{\'e}ger},
  {Sellgren}, \& {Schmidt}}]{Moutou2000}
{Moutou}, C., {Verstraete}, L., {L{\'e}ger}, A., {Sellgren}, K., \& {Schmidt},
  W. 2000, \aap, 354, L17.
\newblock \doarXiv{astro-ph/9912559}

\bibitem[{{Niemann} {et~al.}(2010){Niemann}, {Atreya}, {Demick}, {Gautier},
  {Haberman}, {Harpold}, {Kasprzak}, {Lunine}, {Owen}, \&
  {Raulin}}]{Niemann2010}
{Niemann}, H.~B., {Atreya}, S.~K., {Demick}, J.~E., {et~al.} 2010, Journal of
  Geophysical Research (Planets), 115, E12006, \dodoi{10.1029/2010JE003659}

\bibitem[{{Nixon} {et~al.}(2021){Nixon}, {Irwin}, {Sung}, \&
  {Teanby}}]{nixon2021}
{Nixon}, C., {Irwin}, P., {Sung}, K., \& {Teanby}, N. 2021, {Trace gases in
  Titan's Atmosphere with JWST MIRI}, JWST Proposal. Cycle 1, ID. \#2524

\bibitem[{Nixon {et~al.}(2009)Nixon, Jennings, Flaud, Bézard, Teanby, Irwin,
  Ansty, Coustenis, Vinatier, \& Flasar}]{Nixon2009}
Nixon, C., Jennings, D., Flaud, J.-M., {et~al.} 2009, Planetary and Space
  Science, 57, 1573–1585, \dodoi{10.1016/j.pss.2009.06.021}

\bibitem[{Nixon {et~al.}(2010)Nixon, Achterberg, Teanby, Irwin, Flaud, Kleiner,
  Dehayem-Kamadjeu, Brown, Sams, Bézard, Coustenis, Ansty, Mamoutkine,
  Vinatier, Bjoraker, Jennings, Romani, \& Flasar}]{Nixon2010}
Nixon, C.~A., Achterberg, R.~K., Teanby, N.~A., {et~al.} 2010, Faraday
  Discuss., 147, 65, \dodoi{10.1039/C003771K}

\bibitem[{{Nixon} {et~al.}(2012){Nixon}, {Temelso}, {Vinatier}, {Teanby},
  {B{\'e}zard}, {Achterberg}, {Mandt}, {Sherrill}, {Irwin}, {Jennings},
  {Romani}, {Coustenis}, \& {Flasar}}]{Nixon2012}
{Nixon}, C.~A., {Temelso}, B., {Vinatier}, S., {et~al.} 2012, \apj, 749, 159,
  \dodoi{10.1088/0004-637X/749/2/159}

\bibitem[{Nixon {et~al.}(2013)Nixon, Jennings, Bézard, Vinatier, Teanby, Sung,
  Ansty, Irwin, Gorius, Cottini, \& et~al.}]{Nixon_2013}
Nixon, C.~A., Jennings, D.~E., Bézard, B., {et~al.} 2013, The Astrophysical
  Journal, 776, L14, \dodoi{10.1088/2041-8205/776/1/l14}

\bibitem[{{Nixon} {et~al.}(2016){Nixon}, {Achterberg}, {{\'A}d{\'a}mkovics},
  {B{\'e}zard}, {Bjoraker}, {Cornet}, {Hayes}, {Lellouch}, {Lemmon},
  {L{\'o}pez-Puertas}, {Rodriguez}, {Sotin}, {Teanby}, {Turtle}, \&
  {West}}]{Nixon2016}
{Nixon}, C.~A., {Achterberg}, R.~K., {{\'A}d{\'a}mkovics}, M., {et~al.} 2016,
  \pasp, 128, 018007, \dodoi{10.1088/1538-3873/128/959/018007}

\bibitem[{Nixon {et~al.}(2019)Nixon, Ansty, Lombardo, Bjoraker, Achterberg,
  Annex, Rice, Romani, Jennings, Samuelson, \& et~al.}]{Nixon2019}
Nixon, C.~A., Ansty, T.~M., Lombardo, N.~A., {et~al.} 2019, The Astrophysical
  Journal Supplement Series, 244, 14, \dodoi{10.3847/1538-4365/ab3799}

\bibitem[{Orton {et~al.}(2014{\natexlab{a}})Orton, Fletcher, Moses, Mainzer,
  Hines, Hammel, Martin-Torres, Burgdorf, Merlet, \& Line}]{ORTON2014a}
Orton, G.~S., Fletcher, L.~N., Moses, J.~I., {et~al.} 2014{\natexlab{a}},
  Icarus, 243, 494, \dodoi{https://doi.org/10.1016/j.icarus.2014.07.010}

\bibitem[{Orton {et~al.}(2014{\natexlab{b}})Orton, Moses, Fletcher, Mainzer,
  Hines, Hammel, Martin-Torres, Burgdorf, Merlet, \& Line}]{ORTON2014b}
Orton, G.~S., Moses, J.~I., Fletcher, L.~N., {et~al.} 2014{\natexlab{b}},
  Icarus, 243, 471–493, \dodoi{10.1016/j.icarus.2014.07.012}

\bibitem[{Peeters {et~al.}(2004)Peeters, Mattioda, Hudgins, \&
  Allamandola}]{peeters2004polycyclic}
Peeters, E., Mattioda, A., Hudgins, D., \& Allamandola, L. 2004, The
  Astrophysical Journal, 617, L65

\bibitem[{Perrin {et~al.}(2021)Perrin, Carrasco, Chatain, Jovanovic, Vettier,
  Ruscassier, \& Cernogora}]{perrin2021}
Perrin, Z., Carrasco, N., Chatain, A., {et~al.} 2021, Processes, 9, 965

\bibitem[{Rieke {et~al.}(2004)Rieke, Young, Engelbracht, Kelly, Low, Haller,
  Beeman, Gordon, Stansberry, Misselt, {et~al.}}]{rieke2004}
Rieke, G., Young, E., Engelbracht, C., {et~al.} 2004, The Astrophysical Journal
  Supplement Series, 154, 25

\bibitem[{Rieke {et~al.}(2015)Rieke, Wright, B{\"o}ker, Bouwman, Colina,
  Glasse, Gordon, Greene, G{\"u}del, Henning, {et~al.}}]{rieke2015mid}
Rieke, G.~H., Wright, G., B{\"o}ker, T., {et~al.} 2015, Publications of the
  Astronomical Society of the Pacific, 127, 584

\bibitem[{Rowe-Gurney {et~al.}(2021)Rowe-Gurney, Fletcher, Orton, Roman,
  Mainzer, Moses, de~Pater, \& Irwin}]{Rowe_Gurney_2021}
Rowe-Gurney, N., Fletcher, L.~N., Orton, G.~S., {et~al.} 2021, Icarus, 114506,
  \dodoi{10.1016/j.icarus.2021.114506}

\bibitem[{Sittler {et~al.}(2020)Sittler, Cooper, Sturner, \& Ali}]{SITTLER2020}
Sittler, E.~C., Cooper, J.~F., Sturner, S.~J., \& Ali, A. 2020, Icarus, 344,
  113246, \dodoi{https://doi.org/10.1016/j.icarus.2019.03.023}

\bibitem[{{Spitzer Science Center}(2020)}]{https://doi.org/10.26131/irsa430}
{Spitzer Science Center}. 2020, Spitzer Level 1 / Basic Calibrated Data,  IPAC,
  \dodoi{10.26131/IRSA430}

\bibitem[{Sung {et~al.}(2013)Sung, Toon, Mantz, \& Smith}]{Sung2013}
Sung, K., Toon, G., Mantz, A., \& Smith, M. 2013, Icarus, 226, 1499,
  \dodoi{10.1016/j.icarus.2013.07.028}

\bibitem[{Sung {et~al.}(2018)Sung, Toon, Drouin, Mantz, \& Smith}]{SUNG2018}
Sung, K., Toon, G.~C., Drouin, B.~J., Mantz, A.~W., \& Smith, M. A.~H. 2018,
  Journal of Quantitative Spectroscopy and Radiative Transfer, 213, 119,
  \dodoi{https://doi.org/10.1016/j.jqsrt.2018.03.011}

\bibitem[{Teanby {et~al.}(2009)Teanby, Irwin, {de Kok}, Jolly, B{\'e}zard,
  Nixon, \& Calcutt}]{Teanby2009}
Teanby, N., Irwin, P., {de Kok}, R., {et~al.} 2009, Icarus, 202, 620,
  \dodoi{10.1016/j.icarus.2009.03.022}

\bibitem[{Teanby {et~al.}(2019)Teanby, {Sylvestre}, {Sharkey}, {Nixon},
  {Vinatier}, \& {Irwin}}]{Teanby2019}
Teanby, N., {Sylvestre}, M., {Sharkey}, J., {et~al.} 2019, \grl, 46, 3079,
  \dodoi{10.1029/2018GL081401}

\bibitem[{Teanby {et~al.}(2006)Teanby, Irwin, {de Kok}, Nixon, Coustenis,
  Bézard, Calcutt, Bowles, Flasar, Fletcher, Howett, \& Taylor}]{Teanby2006}
Teanby, N., Irwin, P., {de Kok}, R., {et~al.} 2006, Icarus, 181, 243,
  \dodoi{https://doi.org/10.1016/j.icarus.2005.11.008}

\bibitem[{Teanby {et~al.}(2007)Teanby, Irwin, {de Kok}, Vinatier, Bézard,
  Nixon, Flasar, Calcutt, Bowles, Fletcher, Howett, \& Taylor}]{Teanby2007}
---. 2007, Icarus, 186, 364,
  \dodoi{https://doi.org/10.1016/j.icarus.2006.09.024}

\bibitem[{Teanby {et~al.}(2008)Teanby, Irwin, {de Kok}, Nixon, Coustenis,
  Royer, Calcutt, Bowles, Fletcher, Howett, \& Taylor}]{TEANBY2008}
---. 2008, Icarus, 193, 595,
  \dodoi{https://doi.org/10.1016/j.icarus.2007.08.017}

\bibitem[{Teanby {et~al.}(2013)Teanby, Irwin, Nixon, Courtin, Swinyard, Moreno,
  Lellouch, Rengel, \& Hartogh}]{Teanby2013}
Teanby, N., Irwin, P., Nixon, C., {et~al.} 2013, Planetary and Space Science,
  75, 136, \dodoi{https://doi.org/10.1016/j.pss.2012.11.008}

\bibitem[{Teanby {et~al.}(2017)Teanby, B{\'e}zard, Vinatier, Sylvestre, Nixon,
  Irwin, De~Kok, Calcutt, \& Flasar}]{teanby2017}
Teanby, N., B{\'e}zard, B., Vinatier, S., {et~al.} 2017, Nature communications,
  8, 1

\bibitem[{Teanby {et~al.}(2018)Teanby, {Cordiner}, {Nixon}, {Irwin},
  {H{\"o}rst}, {Sylvestre}, {Serigano}, {Thelen}, {Richards}, \&
  {Charnley}}]{Teanby2018}
Teanby, N., {Cordiner}, M.~A., {Nixon}, C.~A., {et~al.} 2018, \aj, 155, 251,
  \dodoi{10.3847/1538-3881/aac172}

\bibitem[{Thelen {et~al.}(2019)Thelen, Nixon, Chanover, Cordiner, Molter,
  Teanby, Irwin, Serigano, \& Charnley}]{THELEN2019}
Thelen, A.~E., Nixon, C., Chanover, N., {et~al.} 2019, Icarus, 319, 417,
  \dodoi{https://doi.org/10.1016/j.icarus.2018.09.023}

\bibitem[{{Tomasko} {et~al.}(2008){Tomasko}, {Doose}, {Engel}, {Dafoe}, {West},
  {Lemmon}, {Karkoschka}, \& {See}}]{Tomasko2008}
{Tomasko}, M.~G., {Doose}, L., {Engel}, S., {et~al.} 2008, \planss, 56, 669,
  \dodoi{10.1016/j.pss.2007.11.019}

\bibitem[{Vinatier {et~al.}(2012)Vinatier, Rannou, Anderson, Bézard, {de Kok},
  \& Samuelson}]{VINATIER2012}
Vinatier, S., Rannou, P., Anderson, C.~M., {et~al.} 2012, Icarus, 219, 5,
  \dodoi{https://doi.org/10.1016/j.icarus.2012.02.009}

\bibitem[{{Vuitton} {et~al.}(2019){Vuitton}, {Yelle}, {Klippenstein},
  {H{\"o}rst}, \& {Lavvas}}]{Vuitton2019}
{Vuitton}, V., {Yelle}, R.~V., {Klippenstein}, S.~J., {H{\"o}rst}, S.~M., \&
  {Lavvas}, P. 2019, \icarus, 324, 120, \dodoi{10.1016/j.icarus.2018.06.013}

\bibitem[{{Werner} {et~al.}(2004){Werner}, {Roellig}, {Low}, {Rieke}, {Rieke},
  {Hoffmann}, {Young}, {Houck}, {Brandl}, {Fazio}, {Hora}, {Gehrz}, {Helou},
  {Soifer}, {Stauffer}, {Keene}, {Eisenhardt}, {Gallagher}, {Gautier}, {Irace},
  {Lawrence}, {Simmons}, {Van Cleve}, {Jura}, {Wright}, \&
  {Cruikshank}}]{Werner2004}
{Werner}, M.~W., {Roellig}, T.~L., {Low}, F.~J., {et~al.} 2004, \apjs, 154, 1,
  \dodoi{10.1086/422992}

\end{thebibliography}

\end{document}